\documentclass[lettersize,journal]{IEEEtran}
\usepackage{amsfonts}
\usepackage[noend]{algpseudocode}
\usepackage{algorithmicx}
\usepackage{algorithm}
\usepackage{array}
\usepackage[caption=false,font=normalsize,labelfont=sf,textfont=sf]{subfig}
\usepackage{textcomp}
\usepackage{stfloats}
\usepackage{verbatim}
\usepackage{graphicx}
\usepackage{cite}
\usepackage[dvipsnames]{xcolor}
\usepackage{amsmath,amssymb,amsthm}
\usepackage[margin=1in]{geometry}
\usepackage{wrapfig}
\usepackage{url}
\usepackage{marvosym}
\usepackage{enumerate}
\usepackage[font=small,labelfont=bf]{caption} 
\usepackage{tikz}
\usepackage{tkz-fct}
\usepackage{hyperref} 
\usepackage{fancyhdr} 
\usepackage{lastpage} 
\usepackage{extramarks} 
\usepackage{listings} 
\usepackage{courier} 
\usepackage{lipsum} 
\usepackage{booktabs}
\usepackage{bm}
\usepackage{setspace}
\usepackage[noend]{algpseudocode}
\usepackage{authoraftertitle}
\usepackage{parallel}
\usepackage{enumerate}
\usepackage{multicol}
\usepackage{tabu}
\usepackage{colortbl}
\usepackage{rotating}
\usepackage{multirow}
\usepackage{makecell}
\usepackage{bbm}
\usepackage[utf8]{inputenc}
\usepackage[english]{babel}
\usepackage{mathtools}
\usepackage{subfig}
\usepackage{siunitx}
\usepackage{amssymb}

\DeclareSymbolFont{Greekletters}{OT1}{cmbr}{m}{n}
\DeclareMathSymbol{\Th}{\mathord}{Greekletters}{"02}
\newcommand{\dd}{\mathrm{d}}
\newenvironment{spmatrix}[1]
{\def\mysubscript{#1}\mathop\bgroup\begin{pmatrix}}
	{\end{pmatrix}\egroup_{\textstyle\mathstrut\mysubscript}}

\newcommand{\eq}[2]{\begin{equation}\label{#1} #2 \end{equation}}
\newcommand{\al}[1]{\begin{align*} #1 \end{align*}}

\newcommand{\bmx}[1]{\begin{bmatrix} #1 \end{bmatrix}}  

\newcommand{\rf}[1]{\eqref{#1}}
\newcommand{\eqs}[2]{\begin{equation} \label{#1}\begin{split} #2\end{split} \end{equation}}

\newcommand{\norm}[1]{\left\lVert#1\right\rVert}

\DeclareMathOperator*{\argmin}{\arg\!\min}

\renewcommand{\a}{{\alpha}}

\newcommand{\y}{\mathbf{y} }
\newcommand{\X}{\mathbf{X} }
\newcommand{\T}{\Theta }
\newcommand{\w}{\mathbf{w}}
\newcommand{\W}{\mathbf{W}}
\newcommand{\M}{\mathbf{M}}
\newcommand{\ee}{\mathbf{e}}

\newcommand{\ve}{\varepsilon}
\newcommand{\tr}{\nabla}
\newcommand{\td}{\tilde}
\newcommand{\g}{\gamma }

\newcommand{\s}{\sigma }
\renewcommand{\b}{\beta}
\renewcommand{\k}{\kappa}
\newcommand{\D}{\Delta}

\renewcommand{\d}{\delta}
\newcommand{\trs}{^\intercal}
\renewcommand{\S}{\Sigma}

\renewcommand{\l}{{\delta}}

\newcommand{\E}{\mathbb{E}}
\newcommand{\var}{\mathrm{Var}}

\newcommand{\R}{\mathbb{R}}
\renewcommand{\L}{\Lambda}
\renewcommand{\O}{\Omega}
\newcommand{\oi}{\O^{-1}}
\newcommand{\trc}[1]{\text{tr}{(#1)}}

\newcommand{\inv}{^{-1}}
\newcommand{\xddots}{%
	\raise 4pt \hbox {.}
	\mkern 6mu
	\raise 1pt \hbox {.}
	\mkern 6mu
	\raise -2pt \hbox {.}
}

\def\*#1{\bm{#1}}

\newcommand{\gr}

\newtheorem{definition}{Definition}
\newtheorem{lemma}{Lemma}

\newtheorem{theorem}{Theorem}
\newtheorem{corollary}{Corollary}

\begin{document}

\title{Distributed Networked Multi-task Learning}

\author{Lingzhou Hong, Alfredo Garcia, \IEEEmembership{Senior Member, IEEE}
  \thanks{This work was supported in part by the National Science Foundation under Award ECCS-1933878 and in part by the Air Force Office of Scientific Research under Grant 15RT0767. }
  \thanks{Lingzhou Hong and Alfredo Garcia are with the Department of Industrial \& Systems Engineering, Texas A\&M University, College Station, TX 77843 USA (e-mail: \{ hlz, alfredo.garcia\}@tamu.edu). }
  }



\maketitle

\begin{abstract}
We consider a distributed multi-task learning scheme that accounts for multiple linear model estimation tasks with heterogeneous and/or correlated data streams. We assume that nodes can be partitioned into groups corresponding to different learning tasks and communicate according to a directed network topology. Each node estimates a linear model asynchronously and is subject to local (within-group) regularization and global (across groups) regularization terms targeting noise reduction and generalization performance improvement respectively. We provide a finite-time characterization of convergence of the estimators and task relation and illustrate the scheme's general applicability in two examples: random field temperature estimation and modeling student performance from different academic districts.
\end{abstract}


\section{Introduction}

\IEEEPARstart{I}{n} the current age of big data, many applications often face the challenge of processing large and complex datasets, which are usually not available in a single place
but rather distributed across multiple locations. Approaches that require data to be aggregated in a central location may be subject to significant scalability and storage challenges.
In other scenarios, data are scattered across different sites and owned by different individuals or organizations. Data privacy and security requirements make it difficult to merge such data in an easy way. In both contexts,  Distributed Learning (DL)\cite{verbraeken2020survey, ben2019demystifying, tang2020communication} can provide feasible solutions by building high-performance models shared among multiple nodes while maintaining user privacy and data confidentiality.

DL aims to build a collective machine learning model based on the data from multiple computing nodes that can process and store data and are connected via networks. Nodes can utilize neighboring information to improve their own performance: rather than sharing raw data, they only exchange model information such as model parameters or gradients to avoid revealing sensitive information.

Federated learning \cite{li2020review, zhang2021survey} is an example of distributed learning that utilizes a central computing center to maintain and update the global model. 
In this approach, each node is connected with the central node and independently performs model training using its local data, and only shares model updates or gradients with the central server. The central server then combines the local updates to update the global model. While addressing some of the challenges associated with centralized or isolated learning schemes,  frequent communication between the central node and the local nodes may lead to congestion bottlenecks.
The process of collecting and assembling a diverse batch of data points in a central location to update a model may imply significant latency, especially when dealing with high data payloads obtained through heterogeneous and correlated streams. 


In contrast, some other peer-to-peer distributed learning schemes
eliminate the need for a central computing center. This decentralization empowers the system with enhanced scalability, preventing information overflow at a central node.  Additionally, it enables the system to efficiently handle growing data volumes, complex learning tasks, and increasing network demands.

In our previous work \cite{hong2022distributed, garcia2020distributed, hong2020distributed}, we studied peer-to-peer distributed schemes where all nodes share the same learning task (i.e. estimation of a linear model). This paper considers a distributed approach to multi-task learning (MTL)\cite{zhang2021survey}, where multiple learning tasks are jointly undertaken by a network of nodes. 
The relation between learning tasks is not known in advance and must be inferred from the data. 

Many existing MTL methods rely on a central fusion node that updates models in response to locally computed gradient updates  \cite{zhang2014regularization, zhang2021survey,crawshaw2020multi}. 
In MTL, an often-assumed premise is the homogeneity of data and the independence of noise, as noted in  \cite{zhang2021survey}.
When data is correlated and the tasks are closely related, such simplification ignores one source of correlation and can lead to poor model quality. Works \cite{liu2017distributed, smith2017federated, wang2016distributed} propose distributed MTL schemes where a central node updates the global model but lacks a finite-time characterization of convergence. The paper most closely related to ours is \cite{Nassif_2016}, which presents an asynchronous approach for linear multitask problems. However, it does not account for common noise in its formulation, and its convergence analysis primarily focuses on the mean square error of estimations. In contrast, our paper offers finite-time characterizations of the convergence of the estimations and task relation precision matrix.

In this paper, we introduce a Distributed and Asynchronous algorithm for Multi-task Learning (DAMTL) that accounts for 
{\em heterogeneous} and {\em correlated} datasets. We provide a finite-time characterization of convergence.
In the considered architecture, we assume that nodes are connected via a directed graph and can be partitioned into groups corresponding to certain criteria (e.g., distance, similarity). We assume that all nodes within a group have the same learning task and that learning tasks are related according to the Gaussian model whose parameters are unknown.

We formulate the multi-task learning as a bi-level optimization problem, where the outer problem involves estimating the covariance matrix for the Gaussian task relationship model, while the inner problem focuses on estimating linear models.
To solve this problem, we analyze a two-timescale distributed algorithm consisting of stochastic gradient descent (SGD) updates for outer and inner problems.
Each node implements an asynchronous SGD with locally computed gradients and regularization updates for the inner problem. One selected node per group (called {\em messenger}), aside from handling the inner problem,  computes the group precision matrix (the inverse of the covariance matrix) updates for the outer problem.
The nodes communicate in {\em (i)} every node periodically broadcasts its model to other nodes in the same group, and {\em (ii)} messengers periodically exchange group updates with each other.
A key feature is that local updates (in the inner problem) take place at a {\em higher} frequency than global updates (in the outer problem) to reduce communication costs and enhance the system's robustness.
Continuous and asynchronous updating
distinguishes the proposed method from existing ensemble learning
methods that often require a {\em synchronized} model aggregation step.

Estimating the precision matrix (the inverse of the covariance matrix) is a key component in the outer problem. It is 
 challenging due to the curse of dimensionality, and the estimation is sensitive to noise and requires a large number of observations relative to the number of variables.  There are several methods have been proposed to address these challenges and provide efficient estimates \cite{fan2016overview, ledoit2022power}.  We consider a ridge ($L_2$) based estimation of the precision matrix \cite{precision}, which does not require the true (graphical) model to be (extremely) sparse and can be easily implemented via gradient descent.

The paper is structured as follows. In section \ref{data}, we introduce a two-timescale stochastic algorithm for distributed MTL estimation that accounts for heterogeneous and correlated datasets. In section \ref{con}, we provide a finite-time characterization of the convergence of regularity measure that captures the performance gap between the parameter estimation and the ground truth for the inner problem. For the outer problem, we provided the convergence analysis for the regularity measure that describes the distance of the estimated precision matrix to the true task relation precision matrix.
Finally, in section \ref{ex}, we report the method's performance on a synthetic temperature estimation problem and a real dataset on students' study performance.

\section{Data and Processing Model}\label{data}
\subsection{Data Model}

We consider a set of nodes $\mathcal{V} = \{1,\dots, N\}$ with the ability to collect and process data streams $\y_{i}=\{\y_{i,k}\in \R^m|k\in \mathbb{N}^+\}$ of the form: 
\eq{y1}{\y_{i,k}=\X_i\*w_i^*+\ve_{i,k}+\L_i\xi_k, \quad i\in\mathcal{V}}
where  $\X_i\in \R^{m\times p}$ is a random data matrix with rows independent and identically sampled from a multivariate normal distribution $N(\nu_i, \Psi_i)$,  $\*w_i^*\in \R^p$ is the ground truth coefficient vector of coefficients for node $i$. Here $\{\ve_{i,k}\in\R^{m}|k \in \mathbb{N}^+\}$ are the {\em individual noise}, which are
independent and identically distributed (i.i.d.) random noise that only influence node $i$. On the other hand, $\{\xi_k\in\R^{m}|k \in \mathbb{N}^+\}$ are independent realizations of a {\em common noise}  which affects node $i$ according to the matrix $\L_i\in\R^{m\times m}$, thereby introducing correlation across the noise terms from all nodes.
We consider $\L_i$'s as a diagonal matrix with diagonal entries that are possibly different.

We assume individual noise are zero-mean $\E[\ve_{i,k}]=\mathbf{0}_{m \times 1}$ and independent across different nodes, i.e., $\E[\ve_{i,k}\ve_{j,k}\trs]=\mathbf{0}_{m \times m}$ for all $i,j\in \mathcal{V}$, and $j\neq i$. Moreover, we assume 
$\E\norm{\ve_{i,k} \ve_{i,k}\trs}={ \sigma_{i}^2}\mathbf{I}_m$, and $\sigma_i$'s may differ across nodes,  making the noise term heterogeneous.
The common noise vectors are i.i.d with
$\E[\xi_k]=\mathbf{0}_{m}$ and $\E\norm{\xi_k}^2=\mathbf{I}_m$.
It follows the covariance matrix of the error term in \rf{y1} for $\y_{i,k}$ as
\al{\O_{i}:=\E[(\ve_{i,k}+ \L_i\xi_k)(\ve_{i,k}+ \L_i\xi_k)\trs]={ \sigma_{i}^2}\mathbf{I}+\L_i^2.}

\subsection{Task Relationship Model}
 We consider a network structure represented by a graph $\mathcal{G}=(\mathcal{V},\mathcal{E})$, where an edge $(i,j)\in \mathcal{E}$ represents the ability to exchange information between nodes $i$ and $j$. This is represented by the adjacency matrix
 $A \in \mathbb{R}^{N \times N}$ with $a_{i,j}=1$ if $(i,j)\in \mathcal{E}$,  and $a_{i,j}=0$ otherwise.

 We further assume the graph is composed of $q>1$ connected sub-graphs (groups)
$\mathcal{G}=\{\mathcal{G}_1,\dots,\mathcal{G}_q\}$ with a specific structure (see Figure 1(a) below): for each group there is a {\em unique} node say $i^* \in \mathcal{G}_i$ (which we refer to as {\em messenger}) with edges $(i^*,j) \in \mathcal{E}$ to outside nodes $j \in \mathcal{G}_k, k\in\{1,\dots,q\} \backslash \{i\}$.
In words, nodes can communicate {\em within} each subgraph (according to the subgraph topology) while there is a unique node for each subgraph which we refer to as {\em messenger} that can communicate with other {\em messenger} nodes from other subgraphs.
We assume the ground truth coefficients in the data model in \eqref{y1} are the same for members within the same subgraph, i.e., 
\al{\*w^*_i=\*w^*_j ~~~~i, j \in \mathcal{G}_k,~~k\in\{1,\dots,q\}}

 To model the relationships between tasks, let $\mathbf{W}^*=[\*w^*_1, \dots, \*w^*_N] \in \R^{p\times N}$ be the matrix containing the ground truth coefficients for the data model in \eqref{y1}. As in \cite{Yu_2014} we assume ground-truth matrix $\mathbf{W}^*$ follows a matrix-variate normal distribution, i.e. $\W^*\sim MN_{pq}(\M^*, \mathbf{I}_m\otimes\S)$, where $\otimes$ denotes the Kronecker product, $\M^*=[\*m^*,\dots, \*m^*] \in \R^{p\times q}$ is the mean matrix with $\*m^*$ as the shared mean vector across all nodes \cite{Yu_2014}. The estimating the matrix $\S$ will be formulated in terms of the precision matrix $\Th \triangleq\S^{-1}$.


\subsection{A Bi-level Formulation of MTL}

The joint estimation of $\W$ and $\Th$ can be formulated as a bi-level optimization problem. To elucidate this formulation, we begin by describing the inner-level (or lower-level) objective in the form of task-regularized least squares: 
 \eqs{bi2}{l(\Th,\W) = \sum_{i=1}^N [\ell_i(\*w_i)+
 \rho_{r}(\W^{(q(i))})],}
 where 
 $\ell_i(\*w_i)$ represents the local weighted least-squares loss function:
\eqs{li}{\ell_i(\*w_i) \triangleq \mathbb{E}[\frac{1}{2} (\y_{i}-\X_{i}\*w_i)\trs\Omega_i^{-1} (\y_{i}-\X_{i}\*w_i)],}
Here $\W^{(q(i))}$ represents the estimation matrix associated with subgraph $\mathcal{G}_{q(i)}$ corresponding to node $i$.
The second term in \eqref{bi2} serves as a task regularization term:
\eq{pen2}{\rho_r(\W^{(q)})\triangleq\trc{(\W^{(q)}-\M^{(q)})\Th_l (\W^{(q)}-\M^{(q)})\trs}    }
This term is the Mahalanobis distance between subgraph estimates $\W^{(q)}$ and their mean. 

To describe the outer-level (or upper-level) objective, let us define:
\al{\W^c=[\*w_1-\bar{\*m}, \dots, \*w_N-\bar{\*m}]}
where $\bar{\*m}=\frac{1}{N}[\sum_{j=1}^N w_{1j}, \dots, \sum_{j=1}^N w_{pj}]\trs$ for a given estimation matrix $\W$.  Here, the averages are taken across the nodes (the rows of $W$). 
Let $\bm{\omega}_i^c$ be the $i$th row of $\W^c$. According to the task relationship model, $\bm{\omega}^c_i\sim N(\*0,\S)$. Hence, the empirical estimator of the covariance matrix $\S$ is estimated by
\[S(\W)=\frac{1}{p}\sum_{k=1}^p\bm{\omega}^c_k(\bm{\omega}^c_k)\trs. \]
However, when $N>p$,  the covariance matrix $S$ becomes singular, and the precision matrix estimation cannot be accurately estimated. As in \cite{precision}, we employ a maximum likelihood estimation based precision matrix estimation with a ridge ($L_2$) type of penalty:
 \eqs{r3}{f(\Th,\W) \triangleq &\trc{S(\W)\Th}+\frac{b}{2}\trc{(\Th-T)\trs(\Th-T)}\\
 &-\log |\Th|,  }
where $b\in (0, \infty)$ is the parameter that controls the magnitude of the penalty and $T$ is a symmetric positive definite target matrix.
Note that the penalty term amounts to a proper ridge penalty, since 
\[\trc{(\Th-T)\trs(\Th-T)}=\norm{\Th-T}^2_F.  \]

Based upon definitions \eqref{bi2} and \eqref{r3}, the bi-level optimization formulation of MTL is:

\eqs{bi}{
&\min_{\Th} f(\Th, \W^{\star}(\Th)\\
 s.t.\quad &\W^{\star}(\Th) \in \argmin_{\W} l(\Th,\W).
}
We refer to $\min_{\Th}f(\Th, \W^{\star}(\Th))$ as the {\em outer problem}, which aims to estimate the task relationship under the assumption that the models $\W^*(\Th)$ are optimal with respect the regularized least squares objective. Specifically, it seeks to estimate the task precision matrix $\Th$. 
The {\em inner proble}m $\min_{\W} l(\Th,\W)$ centers on estimating $\W$ given a task relationship model given by $\Th$.






\subsection{A Distributed Approach to Solve \eqref{bi}}
To address the bi-level optimization problem outlined above, we
consider a two-timescale distributed algorithm that consists of SGD updates for the regularized versions of the outer and inner problems. In this scheme, each node implements an asynchronous SGD with locally computed gradients and regularization updates for the inner problem.

In addition to implementing updates for the inner problem, the {\em messenger}, also updates the solution for the outer problem. 
The
communication requirements are as follows: (i) every node periodically broadcasts
its model to other nodes in the same group, and (ii)
messengers periodically exchange group updates with
each other.

In the distributed learning process, we aim to obtain the estimation matrix $\W$ and the task precision matrix $\Th$. Instead of transforming all the information across the entire network, nodes periodically send their model estimates to their neighboring nodes, while the designated messengers are responsible for transmitting the group estimations across different groups. However, this inter-group communication occurs at a lower frequency than the local updates within each group.
This communication strategy reduces the amount of information exchanged between nodes and ensures that only necessary updates are shared, and strikes a balance between system consistency and computational efficiency.

  In each group $\mathcal{G}_l$, the outer problem is performed only in the group messenger, while the inner problem is computed asynchronously by every node within the group.
 Now we introduce the learning problems in more detail. 

\subsubsection{At local learner (Parameter estimation)}
In the parameter  estimation approach \rf{bi2}, 
each node $i\in \mathcal{G}_l, l\in \{1,\dots,q\}$ solves the following ``localized" convex optimization problem,
\eq{gl2}{\min_{\*w_i} L_i(\Th,\W) \triangleq \big\{ \ell_i(\*w_i)+\d_1 \rho_{i}(\*w)+ \d_2\rho_{r}(\W^{(l(i))})\big\}.}

In this setting, the system experiences two correlated factors, one resulting from the data noise and another from task correlations. We utilize two penalties and each target at one noise source to maintain node cohesion.
The consensus regularization term is defined as
\eq{pen1}{\rho_i(\*w)=\frac{1}{2}\sum_{j\neq i} 
a_{i,j} \norm{\*w_i-\*w_j}^2,     }
where $a_{i,j}$ indicates whether node $i$ and $j$ are neighbors, and nodes can only be neighbors if they belong to the same group. 
The consensus regularization $\delta_1 \rho_i(\*w)\geq 0$ serves as a measure of similarity among the models within the same group, promoting consistency among models that share the same ground truth.  Specially,  we have $\rho_i(\*w)=0$ if and only if $\*w_j=\*w_i$ for all $j\neq i$ where $a_{i,j}=1$ and $i,j \in  \mathcal{G}_l $.

The task penalty, as introduced in \rf{pen2}, is designed to promote cohesion among different groups (tasks). The parameters $\delta_1 > 0$ and $\delta_2 > 0$ determine the extent to which consensus and task regularizations, respectively, impact the optimization process. 
Adjusting these parameters allows for maintaining consistency within the same group and maintaining the system's robustness.

\subsubsection{At messenger (Task relationship estimation)}

The messenger node from group $l$ is responsible for collecting group model updates  $\{\*w_i\in \mathcal{G}_l\}$ (denoted as $\W_{\mathcal{G}_l}$) and sends it to other messengers in the network. It also updates the estimation matrix $\W^{(l)}$ when receiving updates from other groups. 
It further updates the corresponding precision matrix and then sends $\rho_{r}(\W)$ back to other nodes within its group.   
 Note that all messengers update their group estimation $\W$ and $\Th$ asynchronously.

\subsubsection{Stochastic gradient updates}
 

In our approach, a virtual clock is employed to generate ticks (see \ref{realtime} ), where the time gaps between consecutive ticks are kept very small. This design ensures that no two consecutive updates or communication occur at the same tick. 
If at the $k$th tick, 
 the messenger from group $l$ is updating the precision matrix for its group, it utilizes the following precision matrix gradient estimate:
\eqs{df}{\tr_{\Th} f_{l,k}=S_{l,k}+b_k(\Th_{l,k}-Th_k)-\Th_{l,k}\inv -  \bm{\varsigma}_{l,k}, }
where  $\{ b_k\}$ is a nonnegative decreasing sequence of penalty parameters that converge to zero, and $\bm{\varsigma}_{l,k} \in \R^{N\times N}$ is a noise matrix generated from the gradient estimation process. We assume all elements of the noise matrix are independent, and each column follows $\bm{\varsigma}^i_{l,k} \sim \*N(\*0, \iota_l^2\*I)$ i.i.d.

At the same time, if node $i$ (including the messenger)  collects a data point $\y_{i,k}$ at the $k$th tick, and assuming the processing time is negligible, node $i$ computes the following estimation gradient estimate:
\eqs{gradient}{\nabla \ell_{i,k}=& \X_i\trs\oi_i(\X_{i,k}\*w_{i,k}-\y_{i,k}) \\
=& g_{i,k}-\X_{i,k}\trs\oi_i (\ve_{i,k}+\L_i\xi_k),}
where
$g_{i,k}:=\nabla_{\*w_{i}} \ell_i(\*w_{i,k})=\X_{i,k}\trs\oi_i\X_{i,k}(\*w_{i,k}-\*w_i^*)$ is the ``noise-free" gradient.
At the $t$th tick, for group $\mathcal{G}_l$, the basic two-timescale system stochastic gradient update is as follows:
\begin{subequations}
\begin{align}
    \Th_{l,k+1}=&\Th_{l,k}-\b_k\mathbf{1}_{l,k}^{c}\tr_{\Th} f_{l,k},    \quad  \text{(messenger)}   \label{two1} \\
   \*w _{i,k+1}=&\*w _{i,k}-\g[\mathbf{1}^{g}_{i,k}\tr\ell_{i,k}
+ \d_1 \mathbf{1}^{n}_{i,k}\nabla_{\*w_{i}} \rho_i(\*w_k) \nonumber\\&
+\d_2 \mathbf{1}^{r}_{i,k}\nabla_{\*w_{i}} \rho_{r}(\W_{k}^{(l)})],  \quad i\in \mathcal{G}_l \label{two2}
\end{align}
\end{subequations}
where $\nabla_{\*w_{i}} \rho_i(\*w_k)$ is the in-group consensus penalty gradient for node $i$,  and $\nabla_{\*w_{i}} \rho_{r}(\W_{k}^{(l)})$ is the task relationship penalty gradient for node $i$.
Here, $\mathbf{1}^{c}_{l,k}$ is an indicator variable that determines whether there is a gradient update for the outer problem at tick $k$. Additionally, 
$\mathbf{1}^{g}_{i,k}$ , $\mathbf{1}^{n}_{i,k}$, and $\mathbf{1}^{r}_{i,k}$  are  the indicator random variables to whether a gradient estimate $g_{i,k}$, a local consensus gradient $\nabla_{\*w_{i}} \rho_i(\*w_k)$, and a global task penalty gradient  $\nabla_{\*w_{i}} \rho_{r}(\W_{k}^{(l)})$ are obtained, respectively. The inner problem stepsize $\g$ remains constant throughout the process, while the outer problem stepsize ${\b_k}$ is a nonnegative decreasing sequence.


\subsection{Algorithm Illustration}
We illustrate the DAMTL network in Figure 1(a) with $8$ nodes grouped into $3$ groups, and the information exchange process in Figure 1(b).  
 \begin{figure}[htp]
	\centering	
	\subfloat[]{\label{figur:1}\includegraphics[width=60mm]{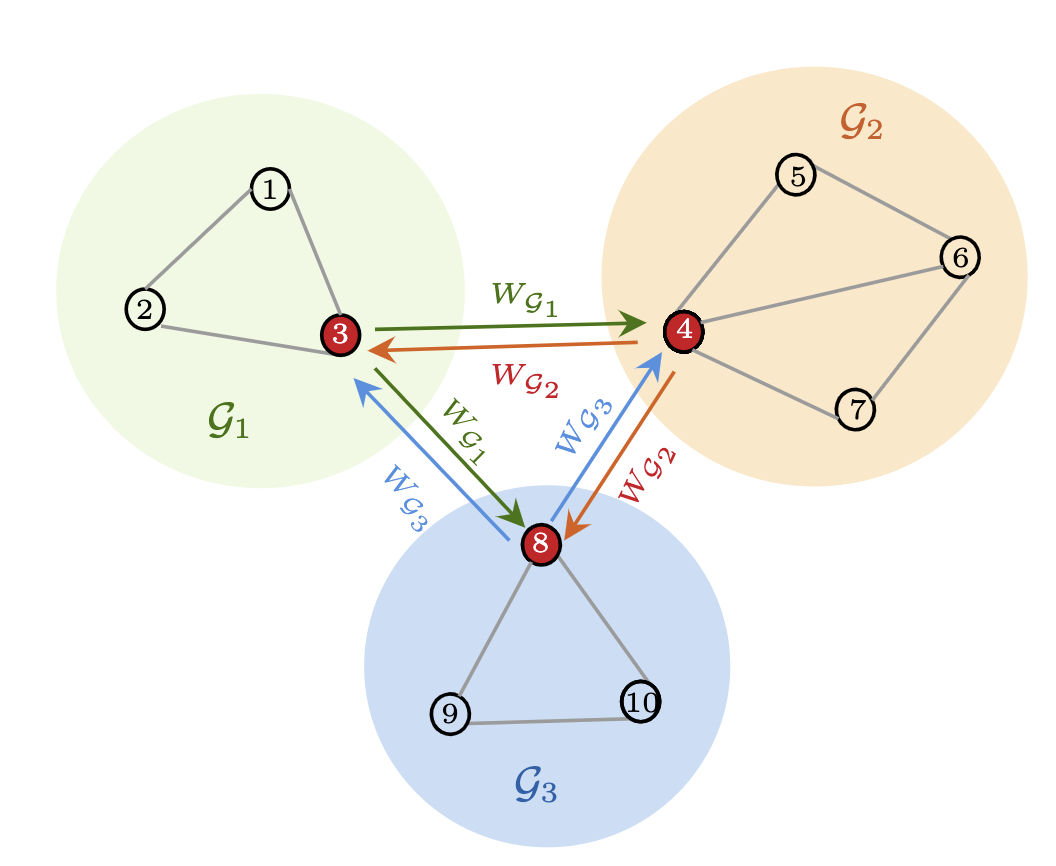}}\\
	\subfloat[]{\label{figur:2}\includegraphics[width=80mm]{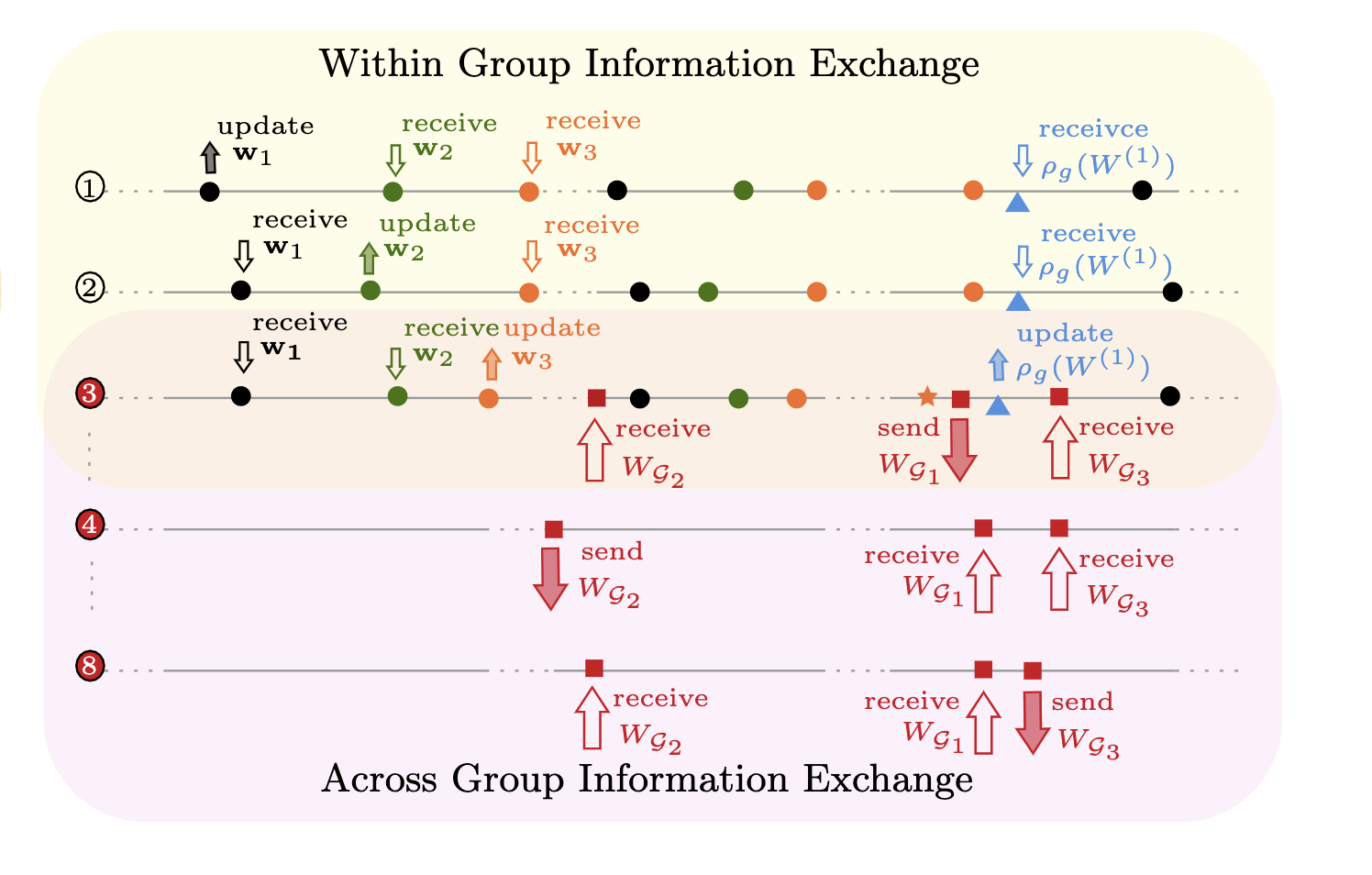}}\\
 	\begin{minipage}{8cm}%
		\small Figure 1. (a) Network structure of DAMTL. There are $8$ nodes assigned into $3$ groups, and nodes $2,4,8$ are messenger nodes of groups $\mathcal{G}_1$, $\mathcal{G}_2$, and $\mathcal{G}_3$ respectively.  (b)  The information exchange process of $\mathcal{G}_1$. The yellow shaded area is the {\em within group} information exchange timeline of $\mathcal{G}_1$, and the red shaded area illustrates the {\em across group} information exchange timeline.
	\end{minipage}%
\end{figure}

In Figure 1(b), when the messenger (node 3) is ready to share the information  (marked by an orange star), it sends out $\W_{\mathcal{G}_1}$ to messengers 4 and 8. At the time to update the estimation matrix (marked by a blue triangle), while $\W_{\mathcal{G}_3}$ has not been received,  the messenger 3 only updates $\W^{(1)}$ with received $\W_{\mathcal{G}_2}$ (during the within-group information exchange, the most recent updates from group members are already stored in $\W^{(1)}$). Subsequently, the messenger 3 sends $\rho_{r}(\W^{(1)})$ to group member 1 and 2.

To illustrate the process, in the following algorithm, we make the assumption that regular nodes update constantly, whereas the messenger node, apart from conducting its estimation update after completing one round of the inner phase, will also perform group updates. Specifically, Algorithm 1 outlines the DAMTL procedure for regular (non-messenger) nodes, while Algorithm 2 outlines the DAMTL procedure for messenger nodes.
\begin{algorithm}
	\caption{DAMTL regular node }
	\begin{algorithmic}[1]
		\State Loops
		 \State  \quad compute available function and penalty gradients
		 \State   \quad update $\*w_{i,k+1}$ by \rf{two2}
          \State  End of Loops
	\end{algorithmic}
\end{algorithm}


\begin{algorithm}
	\caption{DAMTL messenger node $i\in \mathcal{G}_l$, $l\in \{1,\dots,q\}$}
	\begin{algorithmic}[1]
           \State Outer problem loops:
		\State  \quad Inner problem loops:
		 \State      \quad \quad \quad compute available gradients
		 \State      \quad \quad \quad compute $\*w_{i,k+1}$ by \rf{two2}
	    \State \quad  End of the inner problem loops 
		 \State \quad  update $W_{\mathcal{G}_l}$ and send it to other messengers
		 \State\quad update $W^{(l)}$ with collected $W_{\mathcal{G}_j}$'s, $j\in\{1,\dots,q\}$
		 \State \quad  update $S_{l,k}$ and compute $\Th_l$ 
		 \State \quad  update $\rho_{r}(W^{(l)})$ and send it to other nodes in $\mathcal{G}_l$ 
		 \State \quad   update $\tr_{\*w_i} \rho_{r}(W^{(j)})$ 
		 \State End of the outer problem loops
	\end{algorithmic}
\end{algorithm}

\subsection{Continuous Time Approximation}\label{realtime}

For analysis purposes, we embed the discrete-time process into a continuous time setting.
Let the $k$th tick of the global clock happens at $t_k$, for the notation simplicity, we assume $t_k=k$.
For the outer problem, let $N_{c,l}(t)=\sum_{t_k\leq t}\*1_{l,k}^c$ to count for the number of the task precision matrix update for group $l$ that occurred up to time $t$.
We assume that at $k$th tick the messenger of $\mathcal{G}_l$ obtains the $j$th gradient update $\nabla  f_{l, (j)}:=\nabla f_{l,k} $, and denote this time point as $t_{c,l}^{(j)}$. We have the following relationship: $t_{c,l}^{(j)}=k$, $\mathbf{1}^{c}_{l,k}=1$, and $N_{c,l}(k)=j$. We define the random time variable
$\D t_{c,l}^{(j)}=t_{c,l}^{(j)}-t_{c,l}^{(j-1)}$ to compute $\nabla f_{l,(j)}$. 
We assume $\D t_{c,l}^{(j)}$'s  are i.i.d. with $ \E[\D  t_{c,l}^{(j)}]=\D  t_{c,l}$ and $w.p.1$,
\[\lim_{t\to \infty}\frac{N_{c,l}(t)}{t}=\frac{1}{\D t_{c,l}}:=\varphi_l,\]
where $\varphi_l$ can be seen as the precision gradient update rate.

Similarly, for  the inner problem, we define the counting process $N_{g, i}(t)=\sum_{t_k\leq t}\*1_{i,k}^g$ to count the number of function gradient updates $\tr\ell_{i,k}$ occurred  up to time  $t$. Let $N_{n,i}(t)=\sum_{t_k\leq t}\*1_{i,k}^n$ count for local penalty gradient updates $\nabla_{\*w_{i}} \rho_i(\*w_k)$ and $N_{r,i}(t)=\sum_{t_k\leq t}\*1_{i,k}^r$ count for the task penalty gradient  $\nabla_{\*w_{i}} \rho_{r}(\W_{k}^{(l)})$ updates. 
We assume that at the $k$th tick, node $i$ receives the $j$th function gradient update  $\D \ell_{i,(j)}:=\D \ell_{i,k}$. We denote this time point as  
 $t_{g,i}^{(j)}$ with $t_{g,i}^{(j)}=k$, $\mathbf{1}^{g}_{i,k}=1$, and $N_{g,i}(k)=j$. Similarly, let 
the time point of the $j$th local penalty gradient  $\nabla_{\*w_{i}} \rho_i(\*w_{(j)})$ updates as $t_{n,i}^{(j)}=k$ with $\mathbf{1}^{n}_{i,k}=1$ and $N_{n,i}(k)=j$, and the $j$th task penalty gradient $\nabla_{\*w_{i}} \rho_{r}(\W_{(j)}^{(l)})$ updates at $t_{r,i}^{(j)}=k$ with $\mathbf{1}^{r}_{i,k}=1$ and $N_{r,i}(k)=j$. We define the following random time variables:
let $\D t_{g,i}^{(j)}=t_{g,i}^{(j)}-t_{g,i}^{(j-1)}$ be the random time to compute $\nabla \ell_{i,(j)}$, 

For node i, we define the following random time variables:
let $\D t_{g,i}^{(j)}=t_{g,i}^{(j)}-t_{g,i}^{(j-1)}$ be the random time to compute $\nabla \ell_{i,(j)}$, $\D t_{n, i}^{(j)}=t_{n,i}^{(j)}-t_{n,i}^{(j-1)}$ be the random time to collect updated models from neighbors and compute the local penalty gradient  $\nabla_{\*w_i} \rho_{i}(\*w_{(j)})$, and  $\D t_{r,i}^{(j)}=t_{r,i}^{(j)}-t_{r,i}^{(j-1)}$ be the random time to collect the updated $W^{(l)}$ and compute the task penalty gradient  $\nabla_{\*w} \rho_{r}(W_{(j)}^{(l)})$.

We assume $\D t_{g,i}^{(j)}$'s  are i.i.d. with $ \E[\D  t_{g,i}^{(j)}]=\D  t_{g,i}$,  $\D t_{n,i}^{(j)}$'s are i.i.d with $ \E[\D t_{n,i}^{(j)}]=\D t_{n,i}$, and $\D t_{r,i}^{(j)}$'s are i.i.d with $ \E[\D t_{r,i}^{(j)}]=\D t_{r,i}$. Then $w.p.1$,
\al{&\lim_{t\to \infty}\frac{N_{g,i}(t)}{t}=\frac{1}{\D t_{g,i}}:=\mu_i, \\
&\lim_{t\to \infty}\frac{N_{n,i}(t)}{t}=\frac{1}{\D t_{n,i}}:=\varpi_i, \\
&\lim_{t\to \infty}\frac{N_{r,i}(t)}{t}=\frac{1}{\D t_{r,i}}:=\phi_i.}
Here we can see $\mu_i$ as the function gradient update rate, $\varpi_i$ as the local penalty gradient update rate, and $\phi_i$ as the task penalty gradient update rate. Thus, we have changed the count of individual arrival of updates to update rates.  

We rewrite the stepsize of the outer problem  $\beta_k=\beta\hat{\beta}_k$ as the product of a constant $\beta$ and a decreasing sequence $\{\hat{\beta}_k\}$. We introduce the re-scale process for both problems. 
For the outer problem, we define, we define $\T_{l,t}=\Th_{l,t/\beta}$, and for the inner problem, we update $\w_{i,t}:=\*w_{i,t/\gamma}$. and by Donsker's invariance principle \cite{Donsker}, we approximate the rescaled noise terms as Wiener processes under the limits $\beta\to 0$ and $\g\to 0$, respectively.

For the inner problem, let node $i$ generate data points $(\*X_{i,k},y_{i,k}),~k \in \mathbb{N}^+$ instantaneously with rate $\mu_i>0$. When assuming the time required to compute gradient estimates $g_{i,k}$ locally is negligible compared to t time between model updates, it becomes equivalent to assuming that the gradient update rate at node $i$ is $\mu_i$. Additionally, we consider the rate of parameter exchange with neighbors as $\varpi_i$, and we neglect the computation time for the network penalty gradient. As a result, the update rate for the penalty gradient $\nabla \rho_{i,k}$ can be approximated as $\varpi_i$. Similarly, in the context of the outer problem, we assume negligible computation time and a function gradient update rate of $\varphi_l$.
Subsequently, we establish that the dynamics of $\T_{l,t}$ and $\w_{i,t}$  can be modeled via the stochastic differential equations (see Appendix \rf{cts}):
\begin{subequations}
\begin{align}
    \dd \T_{l,t}=&-\varphi_l \hat{\b}_{t} (S_{l,s}+b_t(\T_{l,s}-T_s)-\T_{l,s}\inv)\dd t +\nonumber\\
    &\hat{\b}_{t} \varkappa_l\dd M_{l,t}  \label{dtwo1}\\
   \dd\w_{i,t}=&-(\mu_i g_{i,t}+ \d_1\varpi_i \nabla \rho_{i,t}+\d_2\phi_i\nabla\rho_{r,t}^{(i,j)})dt+\nonumber\\
   &\tau_i dB_{i,t} + \tau_i \dd W_{i,t}.\label{dtwo2}
\end{align}
\end{subequations}
where $\tau_i=\sqrt{\g\mu_i }$, $\varkappa_j= \iota_j \sqrt{\b \varphi_j} $. The noises are approximated by Brownian terms: $B_{i,t}$ is a general $m$ dimensional Brownian motion with covariance $\Upsilon_i$,
i.e, $B_{i,t}=C_iF_t$ with $B_t$ being a standard $m$ dimensional Brownian motion and $C_i\in \R^{p\times p}$ with $C_i C_i\trs=\Upsilon_i$. 
$W_{i,t}$ is a general $m$ dimensional Brownian motion with covariance $\Xi_i$, i.e, $W_{i,t}=D_iW_t$ with $W_t$ being a standard $m$ dimensional Brownian motion and $D_i\in \R^{p \times p}$ with $D_i D_i\trs=\Xi_i$. Here $M_{j,t}$ is a matrix with each column containing a standard $m$ dimensional Brownian motion.

\section{Convergence Analysis}\label{con}
To characterize convergence, we utilize measures of {\em regularity} and {\em consistency}.
The measure of consistency for the outer problem is defined as:
\eqs{V}{V_k=\frac{1}{2q}\sum_{j=1}^q \norm{\Th_{j,k}-\T_*}_F^2= \frac{1}{2q}\sum_{j=1}^q \sum_{i=1}^N \norm{ \Th_{j,k}^i-\Th^i_*}^2,  }
where $\Th_{j,k}^i$ is the $i$th column of $\Th_{j,k}$ and $\Th^i_*$ is the $i$th column of $\T_*$, the ground true relation precision matrix.
For the inner problem, we define the regularity measure as:
\eqs{U}{U_t=\frac{1}{2}\norm{\W_t-\W^*}_F^2=\frac{1}{2}\sum_{i=1}^N\norm{\w_{i,t}-\w_i^*}^2.}
These two metrics serve to gauge the extent to which the estimations deviate from the actual ground truths of local model parameters and the task relationship.

\subsection{Preliminary}
We use similar definitions and results as in \cite{hong2022distributed} in the convergence analysis. 
 Consider the Laplacian matrix $L=D-A$ of graph ${\mathcal{G}}$, where $D$ is the degree matrix, and $A$ is the adjacency matrix. We denote the second smallest eigenvalue of $L$ as $\lambda_2$.
The continuous-time  gradient function $g_{i,t}$ defined above is a function of $\w_{i,t}$, and in our analysis, we denote $g_{i,t}(\w^*)=\X_i\trs\oi_i\X_i(\w^*-\w^*)$ and note that $g_{i,t}(\w)=0$ for all $i\in \mathcal{V}$ and $t$, we write $g(\w^*)$ to simplify the notation. Similarly, we use $g_i$ to denote $g_{i,t}$ when the property holds for all $t$. We note that the corresponding loss function of $g_i$ (the noise-free version of $f_i$) is strongly convex with $\k_i$.

Let $\w_{i,1}$ and $\w_{i,2}$ be two input vectors taken from the function domain, then 
\eqs{conv}{( &g_i(\w_{i,1})- g_i(\w_{i,2}))\trs(\w_{i,1}-\w_{i,2})\\
\geq &\k_i\norm{\w_{i,1}-\w_{i,2}}^2 \geq \k\norm{\w_{i,1}-\w_{i,2}}^2,
}
for  $\k_i:= 2\lambda_{min}(\X_i\trs\oi_i\X_i)$ and $\k:=\min \k_i$, where $\lambda_{min}(\cdot)$ is the smallest  eigenvalue and hence $g_i$ is strongly convex with $\kappa_i$.

We make a set of assumptions to establish bounds within our framework. Firstly, we assume the true relationship covariance matrix and the precision matrix are bounded. We introduce sequences  $\{P_t\}$ and $\{p_t\}$ such that $0<p_t\leq \lambda_{\min}(\S_t)\leq \lambda_{\max}(\S_t) \leq P_t$ are also bounded, ensuring these sequences remain bounded. 
In addition, also assume that the estimated task precision matrices $\T_{l,t}$, $l\in\{1,\dots,q\}$ are bounded,  and hence 
 the sequence $\{Q_t\}$ that upper bound $\norm{\T_t\inv}_F$ and the induced sequence  $\{O_t \}$ that upper bound  $\norm{\T_t}_F$ are also bounded. 
Lastly, we make the assumption that
 $\{F_t\}$ is a bounded sequence that upper bound $[\E([w_{ij}^c]_t)^4]$ with $[w_{ij}^c]_t$ being the $(i, j)$th element of $\W^c$ at time $t$ of any group.

\subsection{Regularity}

The regularity measure $\{U_t, t\geq 0\}$ captures the distance of local model parameters to the ground truth of the inner problem. The following result provides an upper bound on the expected regularity of the estimates at a given time. 
\begin{theorem}\label{theo_u}
Let $\w_{i,t}$ evolve according to continuous time dynamics \eqref{dtwo2}. Then
\al{\E[U_t]\leq & e^{-ct}U_0+
(1-e^{-ct})\frac{\sum_{i=1}^N\tau_i^2(A_i+G_i)}{c}+\\&\int_{0}^{t} e^{ct}h_2(V_t) U_t \dd t} 
where  $c=2\mu\kappa +2\d_1\lambda_2\varpi$ with $\mu=\max \mu_i$ and $\varpi=\max_i \varpi_i$.  The constants $A_i$ and $G_i$ describe the general Brownian terms and are defined in \rf{A} and \rf{B}. The function $h_2(V_t)$ is defined as
\al{h_2(V_t)\sim\d_2\a\phi\frac{h_1(V_t)\sqrt{ P_t}}{u_t}, }
where $h_1(V_t)$ is a function of $V_t$ to bound $\norm{\Theta_{j,t}}_F$ \rf{h1} for all $j\in\{1,\dots,q\}$,  $u_t$ is a small scalar such that $0<u_t\leq \sqrt{U_t}$, and $\a$ is a constant describing the relationship of estimation variation.

\end{theorem}
\proof
See Appendix \rf{thm1} . 
\endproof
 
We can observe that the expected difference in estimates decreases with growing $\d_1$, which penalizes disagreement with neighbors. Similarly, the larger the algebraic connectivity of the network ($\lambda_2$)  or the strong convexity constant $\kappa$ is, the smaller the expected $U_t$. 

Here the function $h_1$ and $h_2$ are functions of $V_t$ that are associated with the bound of the Frobenius norm of the current precision matrix estimation. The term $\int_{0}^{t} e^{ct}h_2(V_t) U_t \dd t$ shows the influence of the precision matrix estimation on the parameter estimation. We note that the scalar $u_t$ is a lower bound of $\sqrt{U_t}$ at time $t$, we will give an alternative estimation for the bound of $h_2(V_t)$ in Theorem $3$ which does not depend on the selection of $u_t$.

\subsection{Consistency}
The consistency measures $\{V_t, t\geq 0\}$ captures the distance of the estimated task precision matrix to the ground true precision matrix for the outer problem. The following result provides an upper bound on the expected task relation of the estimates at a given time.

\begin{theorem}\label{theo_v}
Let $\T_{l,t}$ evolve according to continuous time dynamics \eqref{dtwo1}. Then
\al{\E[V_t]
=&\varphi\int_0^t \hat{\b}_{s} C_1(s)h(U_s) \dd s + \varphi \int_0^t \hat{\b}_{s} C_2(s)   \dd t \\&+ \frac{\varkappa_j^2 N^2 }{2}  \int_0^t \hat{\b}_{t}^2 \dd t
}
where the function $C_1(t)=O_t+\trc{\T_*}\sqrt{\frac{2}{p}}$ and $C_2(t)=\trc{\T_*}P_t\sqrt{N^2+N\frac{Q_t^2}{p_t^2}}+N$ are related to the magnitude of the precision matrix estimation at $t$. The function $h(U_t)$ is defined as a function of the regularity measure $U_t$,
\al{h(U_t)=N\sqrt{F_t}}
to bound the expected estimation matrix.
Here $\{\hat{\b_t}\}$ is a decreasing sequence s.t. $\b_t=\b\hat{\b}_{l,t}$ with $\b$ a small constant smaller than $1$, and $\hat{\b}_t< abs(\frac{\trc{\T_{l,t}}}{2\trc{\T_{l,t}-T}})$ and $\hat{\b}_t\leq \frac{2m_k}{3N\sqrt{F_t}}$ with $m_t\leq \trc{\T_{l,t}}$ for all $l$. 
\end{theorem}

\proof
See Appendix \rf{thm2}. 
\endproof
We write the stepsize $\b_t=\b\hat{\b}_t$ as a production of a constant step $\b$ and a decreasing sequence $\hat{\b}_t$, and require $\int_0^t\hat{\b}_t$ to converge. The bound on $\hat{\b}_t$ ensures the trace of the precision matrix is positive.
We can observe that the bound on the expected $V_t$ is influenced by the  estimation quality of the precision matrices. Here $h(U_t)$ is a function of the regularity measure and describes the bound of the estimation matrix. The term $\varphi\int_0^t \hat{\b}_{s} C_1(s)h(U_s) \dd s $ shows the influence of the parameter estimation on the precision matrix estimation.

Note that as $t$ increases and the estimation of the precision stabilizes, $C_1(t)$ and $C_2(t)$ can be bounded by constants, and we simplify the theorem in the next section.

\subsection{Two Time-scale Systems}

In this section, we discuss the interaction of the two systems. In Theorem \rf{theo_u} and \rf{theo_v}, we can obtain the upper bound of $\E[U_t]$ and $\E[V_t]$ at time $t$. 
 In this theorem, we will discuss the limiting property of the two systems and utilize universal bounds for the sequences $\{P_t\}$, $\{q_t\}$, $\{Q_t\}$, and $\{O_t\}$ and to provide simplified counterpart of the previous theorems.

\begin{theorem}\label{theo_3}
As $t$ increases and the estimations of the relation precision matrix become more stable, let $P=\max P_t$, $\td{p}=\min p_t$, $Q=\max Q_t$, and $O=\max O_t$. Define 
$c_1=N\sqrt{F}\Big(O+\trc{\T_*}\sqrt{\frac{2}{p}}\Big)$, $c_2=2P\trc{\T_*}\sqrt{N^2+N\frac{Q^2}{\td{p}^2}}+N$, and let $t'$ such that  $\int_{t'}^{\infty}\hat{\b}_{t}< \epsilon$ for a given $\epsilon>0$, then 
\eqs{s1}{
   \lim_{t\to \infty}\E[V_t]&
\leq\frac{\varphi( c_1+c_2)+\varkappa_j^2 N^2}{2}\int_0^{t'} \hat{\b}_{t}   \dd t.  }

In the long run, as the estimation matrix $W_{i,t}$ becomes more stable and $\hat{\b}_t$ decreases,  set $F=\max F_t$ and define $c'=2\mu\kappa +2\d_1\lambda_2-\d_2\a\phi O\sqrt{ P}$. When $\d_2\phi<\frac{2(\mu\kappa+\d_1\lambda_2)}{\a O\sqrt{P}}$, it follows that
\eqs{s2}{
 \lim_{t\to \infty}  \E[U_t]&\leq 
\frac{\sum_{i=1}^N\tau_i^2(A_i+G_i)}{c'}.}

\end{theorem}
\proof
See Appendix \rf{thm3}. 
\endproof

In the long-term perspective, we replace the time-dependent functions $C_1(t)$ and $C_2(t)$ with constants $c_1$ and $c_2$, respectively. This simplification is achieved by adopting universal bounds on the approximations of the precision matrix. Similarly, the functions $h_2(V_t)$ and $h(U_t)$ are substituted with constant bounds, which are included in the constants $c_1$ and $c_2$.
Given that $\int_0^{t'} \hat{\b}_{t} \dd t$ remains bounded, the upper bound of $\E[V_t]$ can be  controlled by the precision matrix update rate $\varphi$ and constant portion of stepsize $\b$. These ensures that the fluctuations of $\E[V_t]$ can be bounded within desired limits, supporting the stability and predictability of the system.

In addition to the influence of the local consensus penalty parameters $\d_1$,  the network connectivity $\lambda_2$, and the convexity constant $\k$, we can observe that the upper bound of $\E[U_t]$ is also influenced by the estimation of the precision matrix and the group penalty $\d_2$. The condition involving $\d_2\phi$ requires the group penalty $\d_2$ to be much smaller than that of in group penalty $\d_1$, and the node gradient updates are faster than that of precision matrix updates. This essentially implies that the updates pertaining to the inner problem, which involves parameter estimation, should occur at a much faster rate compared to the updates related to the outer problem, which involves precision matrix estimation. By maintaining  faster updates in the inner problem while maintaining a slower pace in the outer problem, the system attains heightened robustness. 

The following corollary follows immediately from Theorem \ref{theo_3} and provides the choices of the stepsizes and precision update rates to achieve desired bounds on the consistency and regularity bounds.


\begin{corollary}\label{bound}
In the long run, for any positive number $\zeta_1>0$, $\zeta_2>0$, define positive values $\zeta_3$ and $\zeta_4$ such that $\zeta_3+\zeta_4\sim \frac{2\zeta_1}{\td{\b}_{t'}}$
with $\td{\b}_{t'}=\int_0^{t'} \hat{\b}_{t} \dd t $. Adjust the precision matrix update in outer problem as $\varphi\sim \zeta_3\min\{\frac{1}{c_1'+c_2},1 \} $, choose the outer problem constant portion stepsize as $\beta\sim \zeta_4 \min\Big\{\min_j\big[\frac{1}{ N\sqrt{\iota_j \phi_j}}\big],1 \Big\}$, and set the inner problem stepsize as $\g\sim \zeta_2 \min\Big\{ \frac{c'}{\sum_{i=1}^N\mu_i(A_i+G_i)},1 \Big\}$, then we can expect
\[ \E[V_t] \sim \zeta_1, \text{ and }\E[U_t]\sim \zeta_2.\]
\end{corollary}

\proof
See Appendix \rf{lem1} . 
\endproof
The preceding result provides us with choices for the inner and outer problem formulations, as well as the precision matrix updating rate. These choices enable us to attain the predefined upper bounds established for the expected values of $V_t$ and $U_t$, which are set as $\zeta_1$ and $\zeta_2$, respectively.

\section{Numerical Illustration}\label{ex}
In this section, we apply the proposed method to two examples to corroborate the analytical results. First, we apply the DAMTL algorithm to a Gaussian Markov random field (MRF) estimation problem using a wireless sensor network (WSN) with synthetic data to show the effectiveness of the algorithm. Next, we look at a real-world problem: modeling the students' study performance.

In this section, we apply the proposed approach through two illustrative examples. First, we employ the DAMTL algorithm to a Gaussian Markov random field (MRF) estimation problem with a wireless sensor network (WSN) with synthetic data. Next, we explore a tangible real-world challenge involving the modeling of students' study performance.

\subsection{Temperature Estimation of A Field}
In this example, we use a similar setting as \cite{hong2022distributed} 4.1.
We utilize a WSN  for temperature estimation over a 10m$\times$10m field divided into $100$ equal squares.  We assumed that the temperature is uniform within each square, and $N$ sensors are arbitrarily placed on the field. The sensors can be divided into $4$ groups according to geographical location. 
We assume the field's true temperatures are different for the sensors from different groups, and are
 saved in $\R^{100\times 1}$ vectors $\w_i^*$ (note $\w^*_i=\w^*_j$ if $i,j$ in the same group).
The sensors measure the temperature using noisy local observations $\y_i\in \R^{100\times 1}$, which are corrupted by measurement noise $\ve_i$, unique to sensor $i$, and network disturbance $\xi$, shared by all sensors. Each sensor $i$ shares only a portion of $\xi$, based on a matrix $\L_i$ reflecting the sensor's location and the relative distance of the measured square.  We assume $\w^*_i $ is fixed but $\y_i$ changes at each measurement, which can be expressed as follows:
\eqs{ex_y}{\y_i=\w_i^*+\ve_i+\L_i\xi,}
where  $\ve_i\sim N_m(\mathbf{0}, \s_i^2 \mathbf{I})$ and $\xi\sim N_m(\mathbf{0},  \mathbf{I})$. Note that if we set $\X_{i}=\mathbf{I}$,  \rf{y1} and  \rf{ex_y} have the same form. 

Using a Gaussian MRF, we simulate the temperature of the field and allow temperature values to range from $\ang{0}$F to $\ang{255}$F. The field has two heat sources located at $(2m, 8.5m)$ and $(8.5m, 9m)$, and the temperature drops from the heat source at a rate of $\ang{25}$ F/m within a region of influence that spans $5m$ from the source. We connect all nodes in a group with the group messenger and randomly connect nodes to their neighbors within $2.5m$. The messengers from different groups are all connected to transmit group estimations. See Figure 2(a) for sensor locations and the heat map of the field.  Each node has the following local cost as in \rf{li} with $\ell_i(\w_i) \triangleq \frac{1}{2} (\y_{i,k}-\w_i)\trs\Omega_i^{-1} (\y_{i,k}-\w_i)$. The task relation is estimated with \rf{r3} with $T$ as the identity matrix.

\begin{figure}[htp]
	\centering
	\subfloat[]{\label{figur:1}\includegraphics[width=55mm]{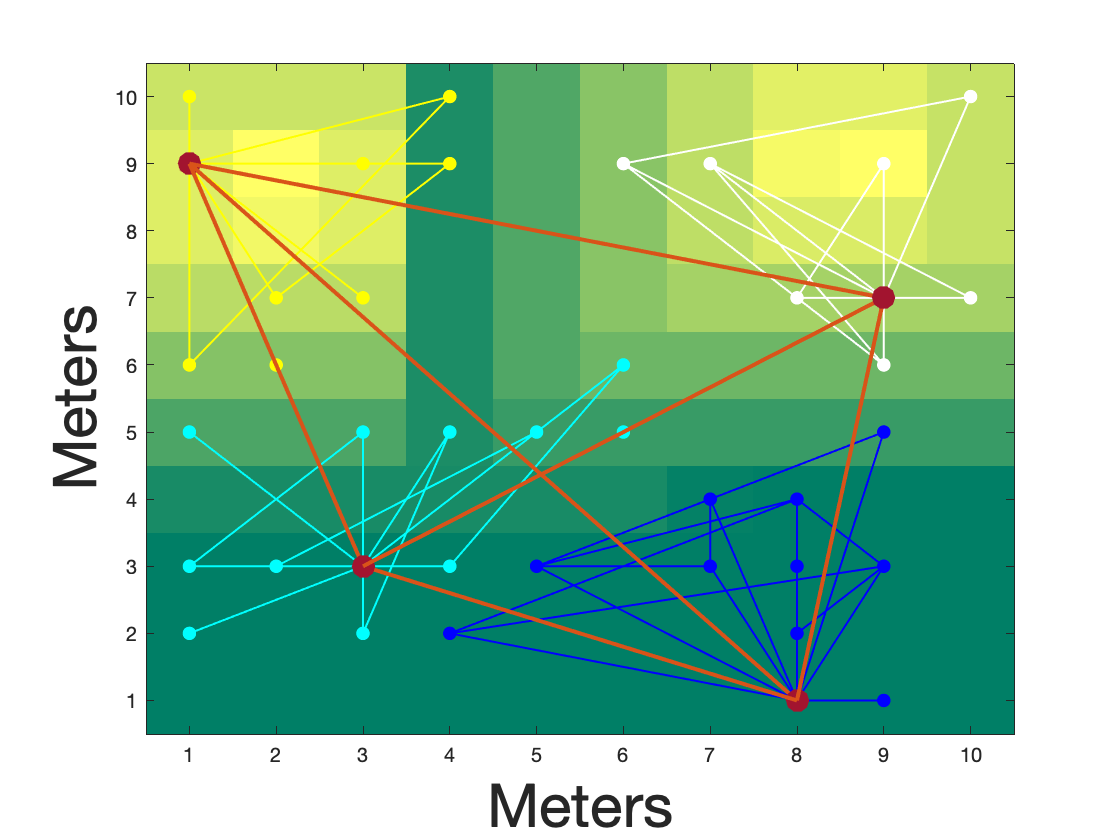}}\\
	\subfloat[]{\label{figur:2}\includegraphics[width=55mm]{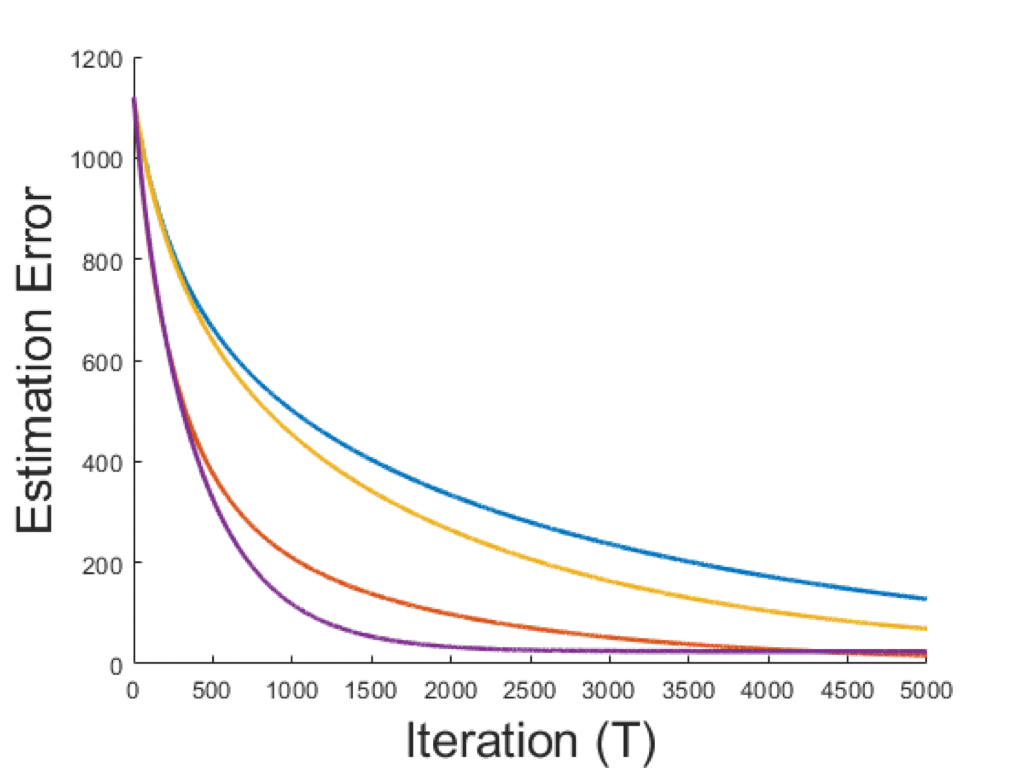}}\\
	\begin{minipage}{8cm}%
		\small Figure 2. (a) Network structure of sensors. The dots denote the nodes and the nodes from the same group are of the same color,
  the  lines represent the edges between nodes. There are $4$ groups and the messenger nodes are marked by red dots. The two heat sources are located at $(2m,8.5m)$ and $(8.5m, 9m)$ and are marked yellow.  (b)  The estimation error comparison of DAMTL, SG, and SG with a partial penalty.
		
	\end{minipage}%
	
\end{figure}

We aim to minimize each sensor's cost function using DAMTL by selecting proper $\w_i$. We use a stepsize of $\g=10^{-4}$ for the inner problem and set the penalty parameter as $\d_1=1500$ and $\d_2=0.9$. For the outer problem,  we utilize a decreasing stepsize $\b_k=\frac{1}{k}$, where $k$ is the iteration. 
We set a minimum individual noise variance of $0.01$, and allow the maximum variance to vary between $1$ and $5$. 
We define the estimation error at time point $k$ as 
\al{Est_k=\frac{1}{4}\sum_{l=1}^4\norm{\W_k^{(l)}-\W^*}_F,}
where $\W_k^{(l)}$ is the estimated matrix at  time point $k$ from group $l$. 
Figure 2(b) compares the estimation error from DAMTL (purple), SG with only task penalty (red), SG with noise penalty (yellow), and SG (blue).  We observe DAMTL has the fastest convergence, which shows its effectiveness. When only using one penalty, the algorithm is not as efficient as DAMTL, which shows the effectiveness of the penalties targeting the noise from both data and task relations.

\subsection{Students' Study Performance}

In this section, we consider a real data set “Junior School Project” from Peter Mortimer \cite{mortimore1989study}, which is a longitudinal study of about $924$ pupils from 50 primary schools chosen at random among the 636 schools under the Inner London Education Authority (ILEA) in 1980. We build a regression model to find the relationship between student scores and other qualifications.

In the model, the $score_5$ is considered as the response variable, and the predictors are {\em gender} (student gender changed to numeral), {\em social} (student father's class, categorical), {\em raven} (raven test score), {\em english} (English test score), {\em math} (math test score).  We normalize all variables and see each school as a node. Though students were tested on the same measure, because of the teaching quality difference, we assume the true model for each school is different yet correlated.
We group all $46$ schools into $5$ groups and select one school as the messenger, we connect all the messengers and randomly connect schools with their neighbors (See Figure 3(a)).
 At each node, the  $score_5$ estimate is given by 
\al{\hat{score_5}=&w_0+w_1\cdot\text{gender}+w_2\cdot\text{social}+w_3\cdot \text{raven}\\&+w_4\cdot\text{englisch}+w_5\cdot\text{math},}

We define the estimation error at time point $k$ for school $i$ as 
\[Err_k(i)=\norm{score_5-\X_i\w_{i,k}}, \]
where $\X_i$ is the matrix containing the observations and $\w_{i,k}$ is the DAMTL estimation for school $i$ at time $k$.  We set the parameter stepsize of the inner problem as $0.0005$ and a decreasing sequence $\b_k=\frac{1}{k}$ for the outer problem. The penalty as $\d_1=20$ and $\d_2=2$.
 Figure 3(b) shows the estimation error of DAMTL for all groups at the final iteration of $50$.
 \begin{figure}[htp]
     	\centering
    	\subfloat[]{\label{figur:2}\includegraphics[width=55mm]{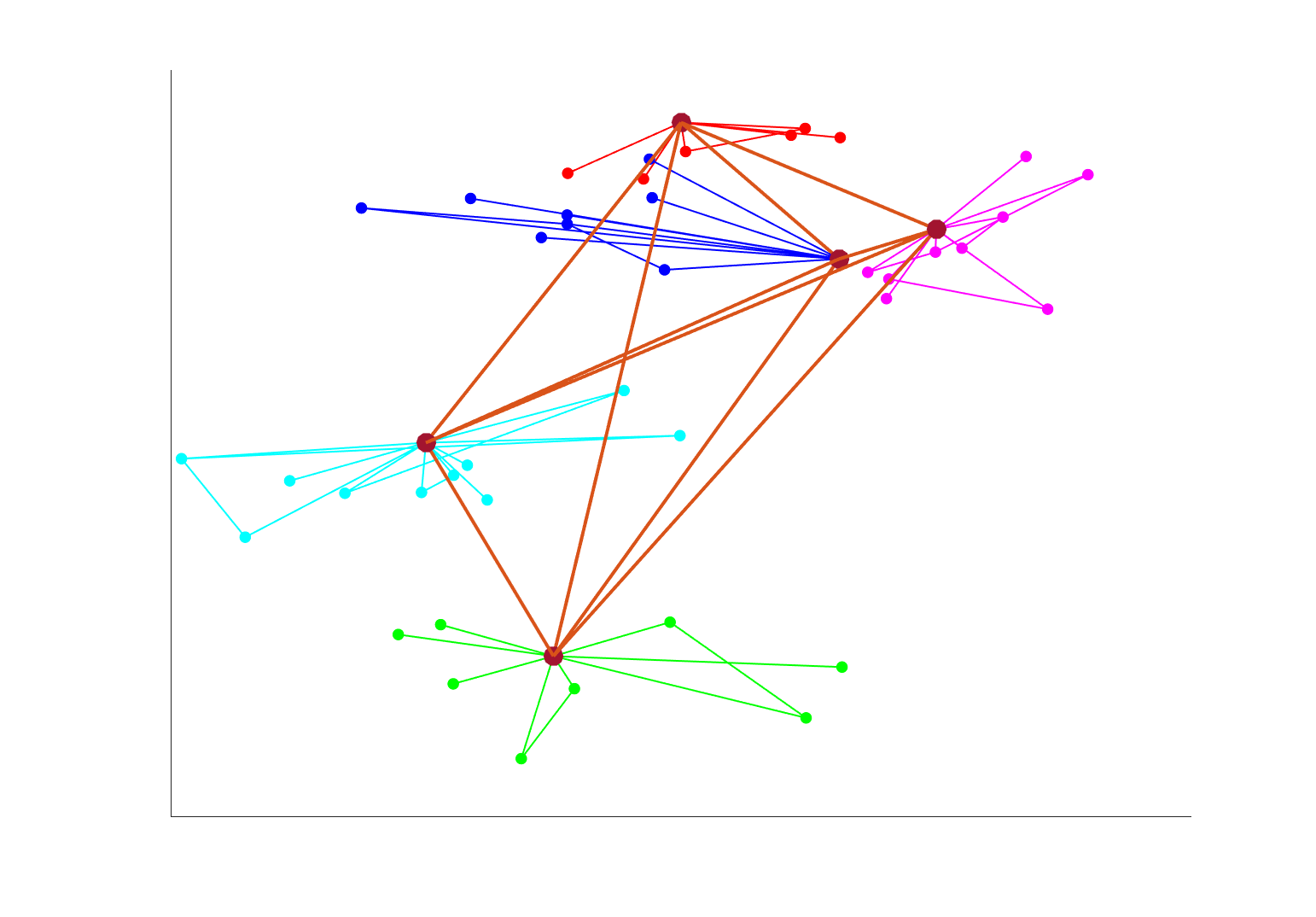}}\\
        \subfloat[]{\label{figur:2}\includegraphics[width=55mm]{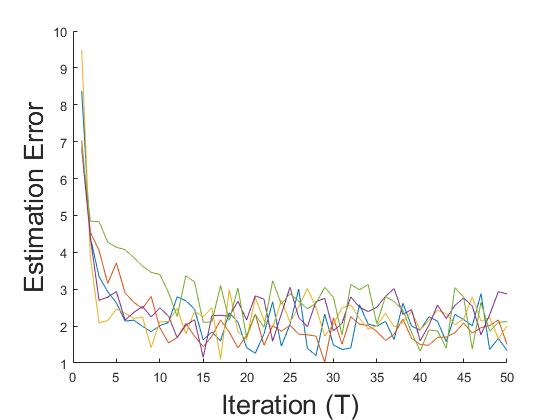}}\\
        
    	\begin{minipage}{8cm}%
		\small Figure 3. (a) network of schools. Schools are denoted by dots and grouped in different colors. Neighboring schools are connected by lines and the selected messengers are denoted by red dots and interconnected.  (b) Estimation error of all groups, and each denoted by a different color.
\end{minipage}%
\end{figure}

The results unveil a rapid convergence of estimations across all schools. The collective average estimation errors across the five distinct groups are showed in the vector $Err=[2.0107,2.0771, 1.7574,2.2998,2.5251]$. We can observe the average error magnitude for each group is around the value of $2$, and the disparities in the error values could potentially result from variations in school characteristics, as well as differences in the quality of teaching provided across these schools.

\section{Conclusion}
The ever-increasing dimension of data and the size of datasets have introduced new challenges to centralized estimation, especially when data comes from streams and the underlying data models are different. In such cases,  we consider a two-timescale distributed learning architecture of nodes that can be partitioned into several interconnected groups.
 In the proposed scheme, each node (or local learner) receives a data stream and {\em asynchronously} implements stochastic gradient updates, and one selected node per group (called {\em messenger}), in a slower frequency, periodically exchange group parameter estimation and estimate task relationship precision matrix. To ensure robust estimation, a local penalty targeting noise reduction and a global regularization targeting overall performance improvement.
  We provide finite-time performance guarantees on the consistency of the parameter estimation and regularity of the precision matrix estimation. We illustrate the application of the proposed method for temperature estimation in a Markov Random Field with synthetic datasets and a real-world problem of students' study performance.

\section{Appendix}
\subsection{Preliminary}
We use the following definition and Ito's lemma.
Let $f:\mathcal{X} \to \mathcal{S}$ be a function with gradient $\tr f(x)$.
\begin{definition}
		Twice differentiable function $f$ is said to be $\k$-strongly convex, if$$(\tr f(x_1)-\tr f(x_2))\trs(x_1-x_2)\geq \frac{\k}{2}\norm{x_1-x_2}^2$$
	for some $\k>0$ and all $x_1$, $x_2\in \mathcal{X}$. Or equivalently, $\tr^2f(x) \succeq\k I$  for all $x\in \mathcal{X}$, i.e., $a_{min}(\tr^2f(x))\geq \k$.
	where $\tr^2f(x) $ is the Hessian matrix, and $a_{min}(\cdot)$ is the minimum eigenvalue.

\end{definition}

\begin{lemma}{Multidimensional Ito Lemma}\label{ito}\cite{oksendal2003stochastic}\\
	Let $X(t)=[X_{1}(t), \dots, X_{p}(t)]\trs$ be a p-dimensional Ito process with
	 \[\dd X(t)=\mathbf{u}\dd t+V\dd B(t),\]
 where $\mathbf{u}$ is a vector of length $p$,  $V$ is a $p\times p$ matrix, and $B(t)=[B_1(t), \dots, B_p(t)]\trs$ is a (standard) $p$-dimensional  Brownian motions. Let $g(x)$ be a twice differentiable map from $\R^p$ into $\R$. Then the process
	\[Y(t)=g(X(t))\]
	is again an  Ito process with 
	\[\dd Y(t)=\sum_{i=1}^p\frac{\partial g(X)}{\partial x_i}\dd X_i +\frac{1}{2}\sum_{i,j}^p\frac{\partial^2g(X)}{\partial x_i \partial x_j}\dd X_i \dd X_j,\]
	where $\dd X_i\dd X_j$ is computed using rules $\dd t\dd t=\dd t\dd B_i=\dd B_i\dd t=0$, $\dd B_i\dd B_j=1$ if $i=j$ and $0$ otherwise.
\end{lemma}

In our setting, $g(x)=\frac{1}{2}\norm{x}^2$ with $\frac{\partial g}{\partial x_i}=x_i$ and $\frac{\partial^2 g}{\partial x_i^2}=1$,  then the process
\al{\dd Y(t)&=\sum_i^p X_idX_i+\frac{1}{2} \dd X \cdot \dd X\\
&=\sum_i^p X_i(u_i\dd t+\sum_k^pV_{i,k}\dd B_k)+\frac{1}{2} \sum_{i,j}^p\dd X_i\dd X_j\\
&=X\trs \dd X+\frac{1}{2}\*1 \trs \dd X \dd X\trs \*1
}
Let $V_{i\cdot }$ denote the $i$th row of $V$, we can expand $\*1 \trs dX dX\trs \*1$ as
\al{&\*1 \trs \dd X \dd X\trs \*1\\
=&\*1 \trs \bmx{u_1\dd t+V_{1\cdot}\dd B\\
\vdots\\
u_p\dd t+V_{p\cdot}\dd B}\Big[u_1dt+V_{1\cdot}\dd B,\dots, u_pdt+V_{p\cdot}\dd B\Big]  \*1\\
=&\sum_{i}\sum_j\sum_k V_{i,k}V_{j,k}\dd t,\\
}

\subsection{Continuous Time Approximation }\label{cts}
In this section, we will derive the formula of $\dd\w_{i,t}$ and $\T_{l,t}$ by writing the scheme \rf{two1} and \rf{two2} in the form of the summation of previous steps, and approximate the noise terms by standard $m$-dimensional Brownian motions and the rest by integrals. Then  $\dd\w_{i,t}$ and $\dd\T_{l,t}$ can be approximated by the differential form of a stochastic Ito integral.  We  assume the noise terms have zero-mean Gaussian distribution: 
$\xi_k\sim \mathcal{N}_m(\mathbf{0},  \mathbf{I}_m)$ and $\ve_{i,k}\sim \mathcal{N}_m(\mathbf{0}, \s_i^2 \mathbf{I}_m)$ for all $i$ in the inner problem, and for each column of $\varsigma$ in the outer problem $\varsigma^i_{j,k} \sim \*N(\*0, \iota_j^2)$. Let $\*v^{(q)}$ denote the component on the $q$th dimension of $\*v$.
\subsubsection{Inner Problem Approximation}
\quad

For each node $i\in\mathcal{G}_j$ and all $j$, we rewrite the scheme \rf{two2} as:
\eqs{01}{&\w_{i,t}
=\w_{i,0}-\g \sum_{l=1}^{N_{g,i}(t/ \g)}g_{i,(l)}
- \g\d_1 \sum_{l=1}^{N_{n,i}(t/ \g)}\nabla_{\*w_{i}} \rho_{i,(l)}(\*w)\\&
-\g\d_2  \sum_{l=1}^{N_{r,i}(t/ \g)}\nabla_{\*w_{i}} \rho_{r}(\W^{(j)}_{(l)})
+ \g \sum_{l=1}^{N_{g,i}(t/ \g)} \X_{i,(l)}\trs\oi_i\ve_{i,(l)}\\&
+ \g \sum_{l=1}^{N_{g,i}(t/ \g)}\X_{i,(l)}\trs\oi_i\L_i\xi_{(l)}}
where $\nabla \ell_{i,k}= g_{i,k}-\X_{i,k}\trs\oi_i (\ve_{i,k}+\L_i\xi_k)$ with
$g_{i,k}:=\X_{i,k}\trs\oi_i\X_{i,k}(\*w_{i,k}-\*w^*)$.
 
Consider the second to the fourth term in \rf{01}.  In the re-scale process, we “squeeze” $N_{g,i}(t/\g)$ function gradient updates in the interval $[0,t]$. We assume  $\g\ll \mu_i$ and partition this interval into subintervals of length $\frac{\g}{\mu_i}=\g\D t_{g,i}$.
It follows that:
\eqs{02}{  \g\sum_{l=1}^{N_{g,i}(t/\g)} g_{i,(l)}=&\frac{\g N_{g,i}(t/\g)}{t}\sum_{l=1}^{N_{g,i}(t/\g)} g_{i,(l)}\frac{t}{N_{g,i}(t/\g)}\\
\approx&\frac{ N_{g,i}(t/\g)}{t/\g}\sum_{l=1}^{N_{g,i}(t/\g)} g_{i,(l)}\frac{\g}{\mu_i}\\
\approx &\mu_i \int_{0}^{t} g_{i,s} ds.}  

Similarly, we have the continuous approximation for the local penalty
\eqs{03}{\g\sum_{l=1}^{N_{n,i}(t/\g)}\tr\rho_{i,(l)}=&\frac{ N_{n,i}(t/\g)}{t/\g}\sum_{l=1}^{N_{n,i}(t/\g)} \tr\rho_{i,(l)}\frac{t}{N_{n,i}(t/\g)}\\
	 \approx&\varpi_i \int_{0}^{t}\tr\rho_{i,s} ds.}

and the continuous approximation for the task penalty
\eqs{04}{ &\g \sum_{l=1}^{N_{r,i}(t/\g)}\tr_{ \w_i}\rho_{r}(\W^{(j)}_{(l)})\\
=&\frac{ N_{r,i}(t/\g)}{t/\g}\sum_{l=1}^{N_{n,i}(t/\g)} \tr_{\w_i}\rho_{r,(l)}^{(j)}\frac{t}{N_{r,i}(t/\g)}\\
\approx&\phi_i \int_{0}^{t}\tr\rho_{r,s}^{(i,j)} ds.}
Here, $\tr\rho_{r,s}^{(i,j)}$ denotes the  task penalty gradient update  at time $s$ of the $i$th node from the $j$th group.


Now consider the individual noise. We assume $\ve_{i,(l)}\sim \mathcal{N}_m(\mathbf{0}, \s_i^2 \mathbf{I})$,   and all components of $\ve_{i,(l)}$ are independent.
We assume that each row of  $\X_i$  is independent and identically sampled from a multivariate normal distribution $N(\nu_i, \Psi_i)$ and $\L_i$ is diagonal.
Since $\oi_i$ is a diagonal matrix, $\X_i\oi_i$ is a random matrix with rows independent and identically follow $N(\oi_i\nu_i,\oi_i\Psi\oi_i )$. 
Let $\zeta_{i,l}:=\X_{i,(l)}\trs\oi_i\ve_{i,(l)}$, then the $q$th component of  $\zeta_i$ can be seen as a linear combination of the components of $\ve_{i,(l)}$ and the $q$th row of $\X_i\trs\oi_i$:
\al{\zeta_{i,l}^{(q)}=\sum_{k=1}^d(\X_{i,(l)}\trs\oi_{i,(l)})_{q,k}(\ve_{i,(l)})_{k}.}
Note that for $\forall k$, the pair $(\X_{i,(l)}\trs\oi_{i,(l)})_{q,k}$ and $(\ve_{i,(l)})_{k}$ are two independent normal random variables, their product is a linear combination of two degree $1$ Chi-square random variables $\frac{1}{4} \big( (\X_{i,(l)}\trs\oi_{i,(l)})_{q,k}+(\ve_{i,(l)})_{k}   \big)^2 -\frac{1}{4}\big( (\X_{i,(l)}\trs\oi_{i,(l)})_{q,k}-(\ve_{i,(l)})_{k}   \big)^2$, which are non-central and dependent. 
For any two  $j\neq k$,  $(\X_{i,(l)}\trs)_{q,j}$ and  $(\X_{i,(l)}\trs)_{q,k}$ are correlated and hence $(\X_{i,(l)}\trs\oi_{i,(l)})_{q,k}(\ve_{i,(l)})_{k}+(\X_{i,(l)}\trs\oi_{i,(l)})_{q,j}(\ve_{i,(l)})_{j}$ follows gamma or generalized gamma for different values correlation coefficient \cite{sumchi}.  We can see $\zeta_{i,l}^{(q)}$ as a  linear combination of several correlated (generalized) Gamman distributed random variables, according to \cite{sumgamma,chiapproximated}, it can be approximated as
a new (generalized) Gamma random variable. Hence  $\zeta_{i,l}$ can be considered as a multivariate Gamma distributed random vector with correlation \cite{gammamultivariate}.

We assume $\zeta_i's$ are i.i.d with $\E [\zeta_{i,l}^2]=\Upsilon_i$ for all $l \in \mathbb{N}^{+}$, since
\[ \E\Big[\sum_{l=1}^{N_{g,i}(t/ \g)} \zeta_{i,l} \Big]= 
 \sum_{l=1}^{N_{g,i}(t/ \g)} \E[\X_{i,(l)}\trs\oi_i]\E[\ve_{i,(l)}]=0, \]
then by multidimensional Donsker's theorem \cite{multiCLT}, 
\[\frac{\sqrt{t}}{\sqrt{N_{g,i}(t/ \g)}}\sum_{l=1}^{N_{g,i}(t/ \g)} \zeta_{i,l}\xrightarrow{\text{d}} B_{i,t}, \]
where $\xrightarrow{\text{d}}$ denotes converge in distribution,
$B_{i,t}$ is a general $m$ dimensional Brownian motion with covariance $\Upsilon_i$, i.e, $B_{i,t}=C_iB_t$ with $B_t$ being a standard $m$ dimensional Brownian motion and $C_i\in \R^{p\times p}$ with $C_i C_i\trs=\Upsilon_i$. 
Let $\tau_i=\sqrt{\g\mu_i }$, we can approximate the individual noise term as
\eqs{indi}{\g \sum_{l=1}^{N_{g,i}(t/ \g)} \X_{i,(l)}\trs\oi_i\ve_{i,(l)}\approx \sqrt{\g\mu_i }B_{i,t}=\tau_iB_{i,t}.}

Now we consider the common noise.  Let $\vartheta_{i,l}=\X_{i,(l)}\trs\oi_i\L_i\xi_{(l)}$,
 we note that for $\forall k$, the pair $(\X_{i,(l)}\trs\oi_{i,(l)}\L_i)_{q,k}$ and $(\xi_{(l)})_{k}$ are two  independent normal random variables.
Similar to the argument in the individual noise approximation, 
 $\varsigma_{i,l}$ can be considered as a multivariate Gamma distributed random vector with arbitrary correlation.
We assume $\vartheta_i's$ are i.i.d with $\E [\vartheta_{i,l}^2]=\Xi_i$ for all $l \in \mathbb{N}^{+}$, 
it follows that
\[ \E\big[ \sum_{k=1}^{N_{g,i}(t/\g)}\xi_{l}  \Big]=  0,\]
\[\frac{\sqrt{t}}{\sqrt{N_{g,i}(t/ \g)}}\sum_{l=1}^{N_{g,i}(t/ \g)} \vartheta_{i,l}\xrightarrow{\text{d}} W_{i,t}, \]
where $W_{i,t}$ is a general $m$ dimensional Brownian motion with covariance $\Xi_i$, i.e, $W_{i,t}=D_iW_t$ with $W_t$ being a standard $m$ dimensional Brownian motion and $D_i D_i\trs=\Xi_i$. 
We can approximate the common noise term as
\eqs{common}{\g \sum_{l=1}^{N_{g,i}(t/ \g)} \X_{i,(l)}\trs\oi_i\L_i\xi_{(l)}\approx \sqrt{\g\mu_i }W_{i,t}=\tau_iW_{i,t}.}
Substituting \rf{02},  \rf{03} and \rf{04} to the corresponding terms in \rf{two2}, $\w_{i,t}$ approximately satisfies the following stochastic Ito integral:
\al{  \w_{i,t}=&\w_{i,0} -\mu_i \int_{0}^{t} g_{i,s} ds
	- \d_1 \varpi_i \int_{0}^{t}\tr\rho_{i,s} ds	- \\&
	\d_2\phi_i \int_{0}^{t}\tr\rho_{r,s}^{(i,j)} ds+
	\tau_i \int_{0}^{t}  d B_{i,s}  + \tau_i\int_{0}^{t} d W_{i,s} .     }
Taking the derivative of the above equation, we get \rf{approx}. 
\eqs{approx}{
\dd \w_{i,t}=&-(\mu_i g_{i,t}+ \d_1\varpi_i \nabla \rho_{i,t}+\d_2\phi_i\nabla\rho_{r,t}^{(i,j)})\dd t+ \\&
\tau_i \dd B_{i,t} + \tau_i \dd  W_{i,t}.
}
\subsubsection{Outer Problem Approximation} 
\quad 

Let $\b_k=\b\hat{\b}_k$ as product of a constant $\b$ and a decreasing sequence $\hat{\b}_k$.  In the re-scale process, we “squeeze” $N_{c,j}(t/\b)$ function gradient updates in the interval $[0,t]$. We assume  $\b\ll \varphi_i$ and partition this interval into subintervals of length $\frac{\b}{\varphi_i}=\g\D t_{c,j}$.  For each group $j$, rewrite the scheme \rf{two1} as
\eqs{theta}{ \T_{j,t}
=& \T_{j,0}-\b\sum_{k=1}^{N_{c,j}(t/\b)} \hat{\b}_k (S_{j,k}+\l_3(\T_{j,k}-T_k)-\T_{j,k}\inv) \\&+ \b\sum_{k=1}^{N_{c,j}(t/\b)} \hat{\b}_k  \varsigma_{j,k}
}
consider the second term of \rf{theta}, 
\eqs{t1}{ & \b\sum_{k=1}^{N_{c,j}(t/\b)} \hat{\b}_k (S_{j,k}+\l_3(\T_{j,k}-T_k)-\T_{j,k}\inv)\\
\approx&\frac{ N_{c,i}(t/\b)}{t/\b}\sum_{k=1}^{N_{c,i}(t/\b)}\hat{\b}_{k} (S_{j,k}+\l_3(\T_{j,k}-T_k)-\T_{j,k}\inv)\frac{\b}{\varphi_j}\\
= &\varphi_j \int_{0}^{t} \hat{\b}_{s} (S_{j,s}+\l_3(\T_{j,s}-T_s)-\T_{j,s}\inv)\dd s.\\
}
We assume all elements are independent for the noise term, and each column $\varsigma^i_{j,k} \sim N(\*0, \iota_j^2)$. We approximate $\sum_{k=1}^{N_{c,j}(t/\b)} \varsigma^i_{j,k}$ by a m-dimensional standard Brownian motion $M_{j,t,i}$, denoting the Browninan approximation of the $i$th column of the noise matrix from group $j$ at time $t$.
\eqs{t2}{ \b\sum_{k=1}^{N_{c,j}(t/\b)} \varsigma^i_{j,k}\approx \iota_j \sqrt{\b \varphi_j} M_{j,t,i}=\varkappa_j M_{j,t,i}
}
where $\varkappa_j= \iota_j \sqrt{\b \varphi_j}$. Let $M_{j,t}=[M_{j,t,1}, \dots,M_{j,t,N} ]$, 
then we can express \rf{theta} as
\eqs{t3}{
\T_{j,t}=&\T_{j,0}-\varphi_j \int_{0}^{t} \hat{\b}_{s} (S_{j,s}+\l_3(\T_{j,s}-T_s)-\T_{j,s}\inv)\dd s \\&
+ \varkappa_j\int_0^t \hat{\b}_{s} \dd M_{j,s}.
}
Taking derivative of \rf{t3}, we can get \rf{dT}
\eqs{dT}{ \dd \T_{j,t}=&-\varphi_j \hat{\b}_{t} (S_{j,t}+\l_3(\T_{j,t}-T_t)-\T_{j,t}\inv)\dd t  \\&+ \varkappa_j\hat{\b}_{t} \dd M_{j,t} }






\subsection{Proof of Theorem 1}\label{thm1}
In this section, we consider the regularity measure for the inner problem \rf{U}.
\[U_t=\frac{1}{2}\norm{\W_t-\W^*}_F^2=\frac{1}{2}\sum_{i=1}^N\norm{\w_{i,t}-\w_i^*}^2.\]
Let $\ee_{i,t}=\w_{i,k}-\w_i^*$, then we can rewrite $U_t=\frac{1}{2}\sum_i\norm{\*e_{i,t}}^2$.  In the followings, we first consider the term $\dd \frac{1}{2}\norm{\ee_{i,t}}^2$.

Similar to the discussion in Lemma 1, the individual noise Brownian term can be seen as:
\al{&\*1\trs \dd B_{i,t}\dd B_{i,t}\trs \*1=\*1\trs \dd(C_iB_t)\dd (C_iB_t)\trs \*1\\&
=\*1\trs C_i\dd B_t\dd B_t\trs C_i\trs \*1\trs=2A_i\dd t,}
where $B_t$ is a standard Brownian motion with 
\eqs{A}{A_{i}=1/2\sum_{l}\sum_j\sum_k c^i_{lk}c^i_{jk},}
where $c^i_{lk}$ is the $(l,k)$th element of matrix $C_i$.
While for the common noise, let $W_t$ be a standard Brownian motion, then
\al{\*1\trs \dd W_{i,t}\dd W_{i,t}\trs \*1=\*1\trs D_i\dd W_t\dd W_t\trs D_i\trs \*1\trs=2 G_i\dd t,}
where 
\eqs{B}{G_i=1/2\sum_{l}\sum_j\sum_k d^i_{lk}d^i_{jk},} 
where $d^i_{lk}$ is the $(l,k)$th element of matrix $D_i$.
Then the production of the Brownian terms becomes
\eqs{dwdw}{\dd \ee_{i,t}\cdot \dd \ee_{i,t}=&\tau_i^2\dd B_{i,t}\cdot \dd B_{i,t}+\tau_i^2\dd W_{i,t}\cdot \dd W_{i,t}\\=&2\tau_i^2(A_i+G_i)\dd t.
}

Now we apply the Ito's lemma to $\dd \frac{1}{2}\norm{\ee_{i,t}}^2$, by \rf{approx} and \rf{dwdw},
\eqs{du}{\frac{1}{2} \dd \norm{\ee_{i,t}}^2=&\ee_{i,t}\cdot \dd \ee_{i,t}+\frac{1}{2}\dd \ee_{i,t}\cdot \dd \ee_{i,t}\\
=& -\ee_{i,t}\trs(\mu_i g_{i,t}+ \d_1\varpi_i \nabla \rho_{i,t}+\d_2\phi_i\nabla\rho_{r,t}^{(i,j)}) \dd t\\ &+\tau_i\ee_{i,t}\trs \dd B_{i,t} + \tau_i\ee_{i,t}\trs \dd  W_{i,t}
+\tau_i^2(A_i+G_i)\dd t
}
Then it follows that 
\eqs{dU}{\dd U_t=&-\sum_{i=1}^N\ee_{i,t}\trs\mu_ig_i\dd t -\sum_{i=1}^N\ee_{i,t}\trs \d_1\varpi_i \nabla \rho_{i,t} \dd t-\\ &\sum_{i=1}^N\ee_{i,t}\trs \d_2\phi_i\nabla\rho_{r,t}^{(i,j)} \dd t+\sum_{i=1}^N\tau_i\ee_{i,t}\trs \dd B_{i,t} + \\ &\sum_{i=1}^N\tau_i\ee_{i,t}\trs \dd  W_{i,t} +\sum_{i=1}^N\tau_i^2(A_i+G_i)\dd t.
}
Consider the first term of \rf{dU}, let $\mu=\max \mu_i$ and by strong convexity assumption and the gradient at ground truth $g(\w*)$ is zero, 
\eqs{du1}{ &-\sum_{i=1}^N(\w_{i,t}-\w^*_i)\trs\mu_ig_{i,t}\\
\leq &-\mu\sum_{i=1}^N(\w_{i,t}-\w^*_i)\trs( g_{i,t}-g(\w_i^*))\\
\leq &- \mu\kappa \sum_{i=1}^N \norm{\ee_{i,t}}^2\leq -2\mu\kappa U_t.
}
For the second term of  \rf{dU},
define the stacked vector $\ee_t=[\ee_{1,t}^T, \dots,\ee_{N,t}^T ]^T$. Notice that the consensus penalty only works within a group (we assumed that only group members (except messengers) are connected, and $\w_i^*=\w_j^*$ when $a_{i,j}=1$, otherwise, $a_{i,j}=0$), then
\al{&-\sum_{i=1}^N\varpi_i \sum_{j\neq i} a_{i,j}(\w_{i,t}-\w_{j,t})\trs(\w_{i,t}-\w^*_i)\\
=&\sum_{i=1}^N\varpi_i\norm{\*e_{i,t}}^2\sum_{j=1}^Nl_{i,j}-\sum_{i=1}^N\varpi_i\sum_{j=1}^Nl_{i,j}\*e_{j,t}\trs\*e_{i,t}\\
\leq&-\varpi\*e_t\trs \*L \*e_t\leq -\varpi\lambda_2\sum_{i=1}^N\norm{\*e_{i,k}}^2=-\varpi\lambda_2U_t
}
where $\varpi=\max_i \varpi_i$, $\*L=L\otimes I_m$ with $\otimes$ denotes the Kronecker product, and $\lambda_2$ is the second smallest eigenvalue of $L$.
Note that the column sums of $L$ is zero, hence $\sum_{i=1}^N\varpi_i\norm{\*e_{i,t}}^2\sum_{j=1}^Nl_{i,j}=0$.
Then the second term \rf{dU} becomes
\eqs{du2}{&-\d_1\sum_{i=1}^N \b_i\*e_{i,t}\trs \nabla \rho_{i,t} =-\d_1\sum_{i=1}^N\b_i \sum_{j\neq i} a_{i,j}\*e_{i,t}\trs\*e_{i,t}\\
\leq & -\d_1\lambda_2\varpi\sum_{i=1}^N\norm{\*e_{i,t}}^2\leq -2\d_1\lambda_2\varpi U_t.
} 
We consider the third term of \rf{dU}. Let $\rho_{r,t}^{(i,j)}$ denotes the task relation penalty for node $i$ from group $G_j$ at time $t$. Let $\W_t^{(j)}=[w]^{j,t}_{lk}$ denote the estimation matrix for group $j$ at time $t$, $\M_t^{(j)}=[\*m_t^{(j)},\dots,  \*m_t^{(j)}]$ with $\*m_t^{(j)}=[m^{j,t}_1, \dots, m^{j,t}_p]\trs$ is the ensemble average estimation of $\M$ at $t$ of group $j$, and $\Theta_{t,j}=[\varrho]_{hq}^{j,t}$ is task precision matrix estimation  from group $j$ of $\S^{-1}$ at $t$  (using $\W_t^{(j)}$). 
\al{\*v_{i}^{j,t}:=&\nabla_{\w_{i}}\trc{(\W_t^{(j)}-\M_t^{(j)})\Theta_{t,j} (W_t^{(j)}-M_t^{(j)})\trs}\\
=&\nabla_{\w_{i}}\trc{\Theta_{t,j} (\W_t^{(j)}-\M_t^{(j)})\trs(\W_t^{(j)}-\M_t^{(j)})}\\
=&\nabla_{\w_{i}}\sum_{h=1}^{N}\sum_{k=1}^{N}\varrho^{j,t}_{hk}\sum_{l=1}^{p}(w_{lk}^{j,t}-m_l^{j,t})(w_{lh}^{j,t}-m_l^{j,t}).
}
Here $\*v_{i}^{j,t}=2[\sum_{k=1}^{N}\varrho_{ki}(w_{1k}^{j,t}-m_1^{j,t}), \sum_{k=1}^{N}\varrho_{ki}(w_{2k}^{j,t}-m_2^{j,t}), \dots, \sum_{k=1}^{N}\varrho_{ki}(w_{pk}^{j,t}-m_p^{j,t})]$ is the gradient of $\trc{(\W-\M)\S\inv (\W-\M)\trs}$ from group $j$ at time $t$ of node $i$. The $l$th element of $\*v_{i}^{j,t}$ can be seen as the inner product of the $i$th column of $\Theta_{t,j}$ and $l$th row of $(\W_t^{(j)}-\M_t^{(j)})$.

We estimate the $l$th element of the ensemble average as  $m_l^{j,t}=\frac{1}{N}\sum_{i=1}^Nw_{li}^{jt}$. For node $i$, $(w_{1i}^{j,t}-m_l^{j,t})$ describes the distance of the mean and individual estimations, and we assume $(w_{1i}^{j,t}-m_l^{j,t})\leq  \a\sqrt{[s_{ii}]^j_t}$ with $[s_{ii}]_t^j$ being the $(i, i)$th element of the covariance estimate from group j, and $\a\geq 3$ is a sufficient constant such that  the inequality holds almost sure.
We assume $0<p_t\leq \lambda_{\min}(\S_{j,t})\leq \lambda_{\max}(\S_{j,t}) \leq P_t $, where $\{p_t\}$ and $\{P_t\}$ are two nonnegtive sequences bounding the eigenvalues of the covariance matrix estimations of all groups.
Since $\S_t$ is Hermitian, $\lambda_{min}(\S_{j,t})\leq [s_{ii}]_t\leq \lambda_{max}(\S_{j,t})$ for all $i$  by Schur-Horn theorem.

Note the relationship between $\norm{\T_{j,t}}_F$ and the task measure $V_t$ and $\norm{\T_*}_F$,
\al{\norm{\T_{j,t}}_F\propto V_t-\norm{\T_*}_F.}
We set 
\eqs{h1}{h_1(V_t)=N\max(\sum_{q=1}^N\varrho_{qi}^t)}
as a function of $V_t$ to bound $\norm{\Theta_{j,t}}_F$, then
\al{  \norm{\Theta_{j,t}}_F&=\sqrt{\sum_{i=1}^N\sum_{j=1}^N\varrho_{ij}^2} \leq \sqrt{(\sum_{i=1}^N\sum_{j=1}^N\varrho_{ij})^2} \\
=&\sum_{i=1}^N\big(\sum_{j=1}^N\varrho_{ij}^{t}\big)\leq N \max(\sum_{q=1}^N\varrho_{qi}^t) =h_1(V_t).}
Then the bound of the $l$th element of $\*v_{i}^{j,t}$ becomes
\al{  &\sum_{q=1}^{N}\varrho_{qi}^t(w_{lq}^{j,t}-m_l^{j,t})\leq\a\sum_{q=1}^N \varrho_{qi}\sqrt{[s_{ii}]_t} \\ \leq &\a\sqrt{P_t} \sum_{q=1}^N\varrho_{qi}^t\leq \frac{\a h_1(V_t)\sqrt{P_t}}{N}.
}
Since $U_t$ is nonnative, find $u_t$, s.t $0<u_t\leq \sqrt{U_t}$. Let $\phi=\max \phi_i$, then the third term of \rf{dU} gives
\eqs{du3}{
\d_2\sum_{i=1}^N\phi_i\ee_{i,t}\trs\*v_{i,k}^j\leq  &\d_2\a\frac{h_1(V_t)\sqrt{ P_t}}{N}\sum_{i=1}^N\phi_i\ee_{i,t}\trs \*1\\
\leq&\d_2\a\frac{h_1(V_t)\sqrt{ P_t}}{N}\sum_{i=1}^N\phi_i\norm{\ee_{i,t}}_1\\
\leq &\d_2\a\phi\frac{h_1(V_t)\sqrt{ P_t}}{N}
\sum_{i=1}^N\sqrt{N}\norm{\ee_{i,t}}_2 \\
\leq &
\d_2\a\phi\frac{h_1(V_t)\sqrt{ P_t}}{u_t}
U_t.
}
The last inequality is because $\sum_{i=1}^N\norm{\ee_{i,t}}_1\leq \sqrt{N}\sum_{i=1}^N\norm{\ee_{i,t}}_2\leq \sqrt{N} \sum_{i=1}^N\sqrt{N\norm{\ee_{i,t}}_2^2}=N\sqrt{U_t}\leq \frac{NU_t}{u_t}$.
By \rf{dwdw}, \rf{du1} \rf{du2}, and \rf{du3}, we can rewrite \rf{dU} as
\eqs{dUb}{&\dd U_t
\leq -(2\mu\kappa +2\d_1\lambda_2\varpi-\d_2\a\phi\frac{h_1(V_t)\sqrt{ P_t}}{u})U_t\dd t+\\&
+\sum_{i=1}^N\tau_i\ee_{it}\trs dB_{i,t} + \sum_{i=1}^N\tau_i\ee_{it}\trs d W_{i,t}+\sum_{i=1}^N\tau_i^2(A_i+G_i)\dd t
}
Let $c=2\mu\kappa +2\d_1\lambda_2\varpi$ and $h_2(V_t)=\d_2\a\phi\frac{h_1(V_t)\sqrt{ P_t}}{u_t} $, and  consider
consider the derivative of $e^{ct}U_t$,

\eqs{deu}{ \dd(e^{ct}U_t)=&e^{ct}\dd U_t+ce^{ct}U_t\dd t\\
\leq & e^{ct}h_2(V_t) U_t\dd t+
e^{ct}\sum_{i=1}^N\tau_i^2(A_i+G_i)\dd t \\&
+e^{ct}\sum_{i=1}^N\tau_i\ee_{i,t}\trs \dd B_{i,t} + e^{ct}\sum_{i=1}^N\tau_i\ee_{i,t}\trs \dd W_{i,t}\\
}

Define the summation of Ito terms, 
\[ K \dd\td{B}_t=\sum_{i=1}^N K_{1,i}d\td{B}_t,\]
where $K_{i}\dd\td{B}_t=\tau_i\ee_{i,t}\trs \dd B_{i,t} + \tau_i\ee_{i,t}\trs \dd W_{i,t}$.
Integrating both sides of the inequality in \rf{deu},
\eqs{eu}{U_t&\leq U_0+\int_{0}^{t} e^{ct}h_2(V_t) U_t \dd t \\&
+(1-e^{-ct})\frac{\sum_{i=1}^N\tau_i^2(A_i+G_i)}{c}+\int_0^{t}e^{cs}K\dd \td{B}_s
}

Since the stochastic integral is a martingale,
\[ 
\E\Big[\int_{0}^{t}e^{c s}K\dd\td{B}_s\Big]=0.
\]
Take expectation of \rf{eu}, we can obtain an upper bound of the regularity measure \rf{two2}
\eqs{Ut}{\E[U_t]\leq &e^{-ct}U_0+
(1-e^{-ct})\frac{\sum_{i=1}^N\tau_i^2(A_i+G_i)}{c}\\&+\int_{0}^{t} e^{ct}h_2(V_t) U_t \dd t}

\subsection{Proof of Theorem 2}\label{thm2}

In this section, we consider the task relation measure for the outer problem \rf{two1}. Suppose there are $q$ groups, the task relation measure \rf{V} can be written as the summation of norms of the columns,
\[V_k=\frac{1}{2q}\sum_{j=1}^q \norm{\Th_{j,k}-\T_*}_F^2= \frac{1}{2q}\sum_{j=1}^q \sum_{i=1}^N \norm{ \Th_{j,k}^i-\T^i_*}^2,  \]
where $\Th_{j,k}^i$ is the $i$th column of $\Th_{j,k}$ and $\Th^i_*$ is the $i$th column of $\T_*$.


For each group $j$, let $\*c_{j,t,i}=\T^i_{j,t}-\T^i_*$, we consider the derivative of the term  $\frac{1}{2}\norm{\T^i_{j,t}-\T^i_*}^2$. Note that $\*c_{j,t,i}$ can be considered as the $i$th column of $\T_{j,t}-\T_*$.  By \rf{dT}, we have 
\al{ \dd \*c_{j,t,i}=-\varphi_j \hat{\b}_{t} \Big[S_{j,t}+\l_3\T_{j,t}-T_t)-\T_{j,t}\inv\Big]_i\dd t + \hat{\b}_{t}\varkappa_j\dd M_{j,t,i}     }
where $[\cdot]_i$ denotes the $i$th column of the matrix. 
Note that $ \dd \*c_{j,t,i} \cdot  \dd \*c_{j,t,i}=\hat{\b}_{t}^2\varkappa_j^2 N \dd t$, then
\eqs{dv}{\dd \frac{1}{2}\norm{\*c_{j,t,i}}^2=&\*c_{j,t,i} \cdot \dd\*c_{j,t,i}+\frac{1}{2} \dd\*c_{j,t,i}\cdot \dd\*c_{j,t,i}            \\
=&-\varphi_j \hat{\b}_{t} \*c_{j,t,i}\trs\Big[S_{j,k}+b_t(\T_{j,k}-T_k)-\T_{j,k}\inv\Big]_i \dd t \\& + \varkappa_j \*c_{j,t,i}\trs \dd M_{j,t,i} +\frac{\hat{\b}_{t}^2\varkappa_j^2N}{2} \dd t.
}  
The derivative of the measure \rf{V} is the following:
\eqs{dV}{
&\dd V_t= \frac{1}{2q}\sum_{j=1}^q \sum_{i=1}^N \dd \norm{\*c_{j,t,i}}^2\\
=& -\frac{1}{q}\sum_{j=1}^q \sum_{i=1}^N \varphi_j \hat{\b}_{t} \*c_{j,t,i}\trs\Big[S_{j,k}+b_t(\Th_{j,k}-T_k)-\Th_{j,k}\inv\Big]_i \dd t \\
& + \frac{1}{q}\sum_{j=1}^q \sum_{i=1}^N \varkappa_j \*c_{j,t,i}\trs \dd M_{j,t,i} + \frac{1}{q}\sum_{j=1}^q \sum_{i=1}^N\frac{\hat{\b}_{t}^2\varkappa_j^2N}{2} \dd t\\
=& -\frac{ \hat{\b}_{t} }{q}\sum_{j=1}^q \varphi_j \trc{(\Th_{j,k}-\T_*)\trs(S_{j,k}+b_k(\Th_{j,k}-T_k)-\Th_{j,k}\inv)} \dd t \\
 &+\frac{1}{q}  \sum_{j=1}^q \varkappa_j\sum_{i=1}^N  \*c_{j,t,i}\trs \dd M_{j,t,i} + \frac{\hat{\b}_{t}^2\varkappa_j^2N^2}{2} \dd t.
}

Now we discuss some matrix properties for the $j$th group at time $t$. For convenience, we drop the subscription for now. Denote the empirical covariance as $S=[o_{ij}]_{1\leq i,j\leq N}$, and $o_{ij}=\frac{1}{p}\sum_{l=1}^p w_{l,i}^c w_{l,j}^c $, where $w_{l,i}^c$ is the $i$th element of the $l$th row of $\W^c$. Let $F_t$ be a scalar to bound to  $[\E(w_{ij})^4]$ (and hence $\sqrt{F_t}$ bounds $\E (w_{ij}^2)$), where $[w_{ij}]_t$ the elements of $\W_t^c$ (apply to all groups).
Let $\S_t$ be the covariance matrix for $S_{j,t}$ for all groups $j$:
\[\E[S_{j,t}]=\frac{1}{p}\sum_{l=1}^p\E[\*w_l^c(\*w_l^c)\trs]=\S_t,\]
The empirical covariance matrix entries have the following properties:
\al{\var(o_{ij})= & \frac{1}{p} \var(w_{l,i}^c w_{l,j}^c)=\frac{1}{p}\E( w_{l,i}^c w_{l,j}^c)^2\\\leq &\frac{1}{p}[\E( w_{l,i}^c)^4]^{1/2}[\E( w_{l,j}^c)^4]^{1/2}\leq \frac{F_t}{p},
}
and 
\[\E(o_{ij})=s_{ij},\]
where $\S=[s_{ij}]$ (denoted as $\S_t=[s_{ij}]_t$  ). Then $o_{ij}$ is an unbiased estimator of $s_{ij}$ with a variance $O(1/p)$.
Note that 
\[ \E(o_{ij}-s_{ij})^2 =\E(o_{ij}^2)-2s_{ij}\E o_{ij}+s_{ij}^2\leq \frac{F_t}{p}+s_{ij}^2\]
For the trace of $S$,
\[\trc{S}=\frac{1}{p}\sum_{l=1}^p \trc{\w_l^c(\w_l^c)\trs} =\frac{1}{p}\sum_{l=1}^p\norm{\w_l^c}^2 \]
\[\E\trc{S}=\trc{\E(S)}=\frac{1}{p}\sum_{l=1}^p \sum_{i=1}^N\E [(\w_{l,i}^c)^2]\leq N\sqrt{F_t}.   \]
Now we prove $\trc{\T}$ is positive. We first discuss in the context of iterations and then extend to continuous time. Let $\hat{\b}_k< abs(\frac{\trc{\T_k}}{2\trc{\T_k-T}})$ and $\hat{\b}_k\leq \frac{2m_k}{3N\sqrt{F_t}}$ with $m_k\leq \trc{\T_{j,k}}$ for all $j$.
We prove by induction, firstly assuming  $\trc{\T_0}>0$ and $\trc{\Th_k}>0$,  then
\al{\trc{\Th_{k+1}}&=\trc{\Th_k-\b_k\big(S_k+b_k(\Th_k-T)-\Th_k\inv              \big)}\\
&=\trc{\Th_k}-\b_k\big( \trc{S_k}+b_k\trc{\Th_k-T} \big) \\&+\b_k\trc{\Th_k\inv}          \\
&\geq\trc{\Th_k}-\b_k\big( \trc{S_k}+b_k\trc{\Th_k-T} \big)          \\
&\geq\trc{\Th_k}-\frac{3\b_k}{2}\ \trc{S_k}        \\
&>\trc{\Th_k}-\frac{3}{2}\frac{2m_k}{3N\sqrt{F_t}} \trc{S_k}        \\
&\geq\trc{\Th_k}-\frac{\trc{\Th_k}}{\trc{S_k}} \trc{S_k}=0       
}
We then embed the result in continuous time according to \rf{t1}, then we have $\trc{\T_{k,t}}>0$. We finish discussing the basic matrix properties here, and
continue to consider the bound for \rf{dV}.

Now we consider the first term of \rf{dV}, 
note that
\al{\E\norm{S_{j,t}-\S_{t}}_F^2\leq&\sum_{i=1}^N\sum_{j=1}^N[s_{ij}]_t^2+\frac{N^2F_t}{p}\\
=&\norm{\S_t}_F^2+\frac{N^2F_t}{p},
}
and
\al{\norm{\S_{t}-\T_{t}\inv}_F^2=&\norm{\S_t\S_t\inv(\S_t-\T_t\inv) }_F^2\\
\leq&\norm{\S_t}_F^2\norm{\S_t\inv(\S_t-\T_t\inv)}_2^2\\
=&\norm{\S_t}_F^2\norm{I-(\T_t\S_t)\inv)}_F^2.
}
We assume $\norm{\T_t\inv}_F\leq Q_t$ and hence 
\eqs{O}{\norm{\T_k}_F\leq \frac{\sqrt{N}Q_t}{\l_{\min}(\T_t\inv)^2}:=O_t.}
Here $\{Q_t\}$ is a bounded sequence that bounds $\norm{\T_t\inv}_F$ and $\{O_t\}$ can be obtained from $\{Q_t\}$.
\al{\norm{I-(\T_t\S_t)\inv}_F^2&\leq 2\norm{I}_F^2+2\norm{(\T_t\S_t)\inv}_F^2\\
&\leq2N+2\norm{\S_t\inv}_2^2\norm{\T_t\inv}_F^2\\
&\leq2N+2\lambda_{\max}(\S_t\inv)^2Q_t^2\\
&=2(N+  \lambda_{\max}^2(\S_t\inv)Q_t^2   )
}
Then we have the upper bound of the expected term,
\al{&    \E\norm{S_{j,t}-\T_{j,t}\inv}_F^2=\E\norm{S_{j,t}-\S_{t}+\S_{t}-\T_{j,t}\inv}_F^2\\
\leq& 2\E\norm{S_{j,t}-\S_{t}}_F^2+2\norm{\S_{t}-\T_{t}\inv}_F^2\\
\leq&\frac{2N^2F_t}{p}+4\norm{\S_t}_F^2(N+\lambda_{\max}^2(\S_t\inv)Q_t^2  )
}
We assume that there exist a constant $b$, such that $\norm{\T_{j,t}-T}_F \leq b\norm{\T_{j,t}-\T_*}_F$ and $\hat{b}_t=bb_t$. As  $b_t\to 0$ almost surely, 
the first term of \rf{dV} becomes
\al{&-\trc{(\T_{j,t}-\T^*)\trs(S_{j,t}+b_t(\T_{j,t}-T)-\T_{j,t}\inv)}\\
=&-\trc{\T_{j,t}S_{j,t}}+\trc{\T^*(S_{j,t}-\T_{j,t}\inv)}+\trc{I}-\\&b_t\trc{(\T_{j,t}-\T_*)\trs(\T_{j,t}-T)}   \\
\leq&\trc{\T_{j,t}}\trc{S_{j,t}}+\trc{\T^*}\norm{S_{j,t}-\T_{j,t}\inv}_F+N+\\&
b_t\norm{\T_{j,t}-\T_*}_F\norm{\T_{j,t}-T}_F
\\
\leq&\trc{\T_{j,t}}\trc{S_{j,t}}+N +2\hat{b}_t\norm{\T_{j,t}-\T_*}_F^2\\&+\trc{\T^*}\sqrt{\frac{2N^2F_t}{p}+4\norm{\S_t}_F^2(N+\lambda_{\max}^2(\S_t\inv)Q_t^2  )}
}

Tate expectations of the equation above, 
\eqs{m2}{&\trc{\T_{j,t}}\E\trc{S_{j,t}}+N+2\hat{b}_t\norm{\T_{j,t}-\T_*}_F^2+\\&
\trc{\T^*}\sqrt{\frac{2N^2F_t}{p}+4\norm{\S_t}_F^2(N+\lambda_{\max}^2(\S_t\inv)Q_t^2  )}\\
\leq & O_t N\sqrt{F_t}+N+2\hat{b}_t\norm{\T_{j,t}-\T_*}_F^2+\\&\trc{\T^*}\sqrt{\frac{2N^2F_t}{p}+4\norm{\S_t}_F^2(N+\lambda_{\max}^2(\S_t\inv)Q_t^2  )}\\
=&N\Big(\trc{\T^*}\sqrt{\frac{2F_t}{p}+4\norm{\S_t}_F^2(\frac{1}{N}+\frac{\lambda_{\max}^2(\S_t\inv)Q_t^2  }{N^2})}\\&+ O_t \sqrt{F_t}++1
\Big)+2\hat{b}_t\norm{\T_{j,t}-\T_*}_F^2.
}
Note that
\al{&\norm{\S_t}_F^2\lambda_{\max}^2(\S_t\inv)\\=&\trc{\S_t^2}\lambda_{\max}^2(\S_t\inv)
\leq N \frac{\lambda^2_{\max}(\S_t)}{\lambda_{\min}^2(\S_t)}
\leq  N  \frac{P_t^2}{p_t^2}.
}
Let $\norm{W^*}_F=B^*$, we note that $\sqrt{F_t}$ is related to the regularity measure $U_t$:
\al{\E[U_t]=&\frac{1}{2}\E\norm{W_t-W^*}_F^2\\
\leq&\sum_i\sum_j\E[w_{ij}^2]+\sum_i\sum_j\E[{(w^*)}_{ij}^2]\\\propto& N^2\sqrt{F_t}+ B^*.}
Let $h( U_t)=N\sqrt{F_t}$ ($h(U_t)_3^2=N^2 F_t$) and $\varphi=\max(\varphi_j)$, then we have 
\eqs{dV2}{&\E [\dd V_t] 
=\varphi \hat{\b}_{t}\Big( h(U_t)O_t+ \\& \trc{\T^*}\sqrt{\frac{2 h(U_t)^2}{p}+4N^2 P_t^2+4N\frac{P_t^2 Q_t^2}{p_t^2}       } +1  \Big)+\\
&\frac{1}{q}  \sum_{j=1}^q \varkappa_j\sum_{i=1}^N  \*c_{j,t,i}\trs \dd M_{j,t,i} + \frac{\hat{\b}_{t}^2\varkappa_j^2 N^2}{2} \dd t+2q\varphi\hat{\b}_t\hat{b}_t V_t\\
\leq&\varphi \hat{\b}_{t}\Big( C_1(t)h(U_t)+  C_2(t)\Big)+\frac{1}{q}  \sum_{j=1}^q \varkappa_j\sum_{i=1}^N  \*c_{j,t,i}\trs \dd M_{j,t,i} \\&+\frac{\sum_{j=1}^q\hat{\b}_{t}^2\varkappa_j^2 N^2}{2q} \dd t+2q\varphi\hat{\b}_t\hat{b}_t V_t\\
}
where $C_1(t)=O_t+\trc{\T_*}\sqrt{\frac{2}{p}}$  and $C_2(t)=\trc{\T_*}P_t\sqrt{N^2+N\frac{Q_t^2}{p_t^2}}+N$.
As $t$ increase, $q\varphi\hat{\b}_t\hat{b}_t \to 0$ and the term $2q\varphi\hat{\b}_t\hat{b}_t V_t$ vanishes. 
Note that the stochastic integral is a martingale,
\[\E\Big[\int_0^t \frac{\hat{\b}_{t}}{2q}  \sum_{j=1}^q \varkappa_j\sum_{i=1}^N  \*c_{j,t,i}\trs \dd M_{j,t,i}  \Big]=0\]
Then we can obtain the expected bound for measure \rf{V}:
\eqs{V0}{\E[V_t]&
=\varphi\int_0^t \hat{\b}_{s} C_1(s)h(U_s) \dd s + \\&\varphi \int_0^t \hat{\b}_{s} C_2(s)   \dd t +\frac{\varkappa_j^2 N^2 }{2}  \int_0^t \hat{\b}_{t}^2 \dd t
}

\subsection{Proof of Theorem 3}\label{thm3}
In this section, we will discuss the limiting properties of the two systems.
First, we consider the effects of changing the expression of the bounding sequences related to precision matrices $\T_{j,t}$, then we further consider using the bounding for the estimation matrix $W_{j,t}$ to deliver the long term two system expression.

When $t$ increases, the estimations of the relation precision matrix become more stable, and as the bounding sequences $\{P_t\}$, $\{q_t\}$, $\{Q_t\}$, and $\{O_t\}$ are upper bounded, let $P=\max P_t$, $\td{p}=\min p_t$, $Q=\max Q_t$, and $O=\max O_t$. We can change the function $C_1(t)$ and $C_2(t)$  in \rf{V0} to two constant terms
$c_1=O+\trc{\T_*}\sqrt{\frac{2}{p}}$ and $c_2=2P\trc{\T_*}\sqrt{N^2+N\frac{Q^2}{\td{p}^2}}+N$ respectively.
Then we can rewrite the bound of the two systems \rf{Ut}  and \rf{V0} as
\al{
    \E[V_t]&
=\varphi c_1\int_0^t \hat{\b}_{t}h(U_s) \dd t + \frac{\varphi c_2+\varkappa_j^2 N^2}{2} \int_0^t \hat{\b}_{t}   \dd t \\
   \E[U_t]&\leq e^{-ct}U_0+
(1-e^{-ct})\frac{\sum_{i=1}^N\tau_i^2(A_i+G_i)}{c}\\&+\int_{0}^{t} e^{ct}h_2(V_t) U_t \dd t.
}
where $h( U_t)=N\sqrt{F_t}$, $h_2(V_t)=\d_2\a\phi\frac{h_1(V_t)\sqrt{ P}}{u_t} $, and $ h_1(V_t)=N\max(\sum_{q=1}^N\varrho_{qi}^t)$.

 In the long run, as the estimation matrix $W_{i,t}$ becomes more stable and $\hat{\b}$ decreases, the two systems also become more stable. We set universal bound $F=\max F_t$, and note $h(U_t)\leq N\sqrt{F}$, $h_1(V_t)\leq O$. As $U_t$ becomes small, the bound for $\sqrt{U_t}$ also bounds $U_t$, and thus $h_2(V_t)\leq \d_2\a\phi O\sqrt{ P}$, we change the constants $c$  to $c'=2\mu\kappa +2\d_1\lambda_2-\d_2\a\phi O\sqrt{ P}$, and $c_1$ to
$c_1'=N\sqrt{F}\Big(O+\trc{\T_*}\sqrt{\frac{2}{p}}\Big)$.  
When $\d_2\phi<\frac{2(\mu\kappa+\d_2\lambda_2)}{\a O\sqrt{P}}$, (
$\d_2\ll \d_1$ and $\d_2\ll \mu$ for simplicity), we have $c'
>0$. Note that $\hat{\b}_t \to 0$, we assume that after $t'$, $\hat{\b}_{t'}$ is negligible,  
The two systems can be written as 
\al{
    \E[V_t]&
=\frac{\varphi( c_1'+c_2)+\varkappa_j^2 N^2}{2}\int_0^{\infty} \hat{\b}_{t}   \dd t  \\
   \E[U_t]&\leq e^{-c't}U_0+
(1-e^{-c't})\frac{\sum_{i=1}^N\tau_i^2(A_i+G_i)}{c'}.
}

As $t\to \infty$, the exponential terms vanish,  we can obtain
\al{
    \E[V_t]&
=\frac{\varphi( c_1'+c_2)+\varkappa_j^2 N^2}{2}\int_0^{t'} \hat{\b}_{t}   \dd t \\
   \E[U_t]&\leq 
\frac{\sum_{i=1}^N\tau_i^2(A_i+G_i)}{c'}.
}

 As $t\to \infty$, the exponential terms vanishes 

\subsection{Proof of Lemma} \label{lem1}
Let $\varphi\sim \zeta_3\min\{\frac{1}{c_1'+c_2},1 \} $,  and $\beta\sim \zeta_4 \min\{\min_j\big[\frac{1}{ N\sqrt{\iota_j \phi_j}}\big],1 \}$, such that 
$\zeta_3+\zeta_4\sim \frac{2\zeta_1}{\td{\b}_{t'}}$,
where $\td{\b}_{t'}=\int_0^{t'} \hat{\b}_{t}   \dd t $. Then 
\al{\E[V_t]\sim \zeta_1.}

When $\g\sim \zeta_2 \min\{ \frac{c'}{\sum_{i=1}^N\mu_i(A_i+G_i)},1 \}$, then 
\al{\E[U_t]\leq \g\frac{\sum_{i=1}^N\mu_i(A_i+G_i)}{c'}\sim \zeta_2,}
that is, no matter whether $\frac{c'}{\sum_{i=1}^N\mu_i(A_i+G_i)}<1$ or not, $\E[U_t]\sim\zeta_2 $.


\bibliography{main}

\begin{thebibliography}{10}
\providecommand{\url}[1]{#1}
\csname url@samestyle\endcsname
\providecommand{\newblock}{\relax}
\providecommand{\bibinfo}[2]{#2}
\providecommand{\BIBentrySTDinterwordspacing}{\spaceskip=0pt\relax}
\providecommand{\BIBentryALTinterwordstretchfactor}{4}
\providecommand{\BIBentryALTinterwordspacing}{\spaceskip=\fontdimen2\font plus
\BIBentryALTinterwordstretchfactor\fontdimen3\font minus \fontdimen4\font\relax}
\providecommand{\BIBforeignlanguage}[2]{{%
\expandafter\ifx\csname l@#1\endcsname\relax
\typeout{** WARNING: IEEEtran.bst: No hyphenation pattern has been}%
\typeout{** loaded for the language `#1'. Using the pattern for}%
\typeout{** the default language instead.}%
\else
\language=\csname l@#1\endcsname
\fi
#2}}
\providecommand{\BIBdecl}{\relax}
\BIBdecl

\bibitem{verbraeken2020survey}
J.~Verbraeken, M.~Wolting, J.~Katzy, J.~Kloppenburg, T.~Verbelen, and J.~S. Rellermeyer, ``A survey on distributed machine learning,'' \emph{Acm computing surveys (csur)}, vol.~53, no.~2, pp. 1--33, 2020.

\bibitem{ben2019demystifying}
T.~Ben-Nun and T.~Hoefler, ``Demystifying parallel and distributed deep learning: An in-depth concurrency analysis,'' \emph{ACM Computing Surveys (CSUR)}, vol.~52, no.~4, pp. 1--43, 2019.

\bibitem{tang2020communication}
Z.~Tang, S.~Shi, X.~Chu, W.~Wang, and B.~Li, ``Communication-efficient distributed deep learning: A comprehensive survey,'' \emph{arXiv preprint arXiv:2003.06307}, 2020.

\bibitem{li2020review}
L.~Li, Y.~Fan, M.~Tse, and K.-Y. Lin, ``A review of applications in federated learning,'' \emph{Computers \& Industrial Engineering}, vol. 149, p. 106854, 2020.

\bibitem{zhang2021survey}
Y.~Zhang and Q.~Yang, ``A survey on multi-task learning,'' \emph{IEEE Transactions on Knowledge and Data Engineering}, vol.~34, no.~12, pp. 5586--5609, 2021.

\bibitem{hong2022distributed}
L.~Hong, A.~Garcia, and C.~Eksin, ``Distributed networked learning with correlated data,'' \emph{Automatica}, vol. 137, p. 110134, 2022.

\bibitem{garcia2020distributed}
A.~Garcia, L.~Wang, J.~Huang, and L.~Hong, ``Distributed networked real-time learning,'' \emph{IEEE Transactions on Control of Network Systems}, vol.~8, no.~1, pp. 28--38, 2020.

\bibitem{hong2020distributed}
L.~Hong, A.~Garcia, and C.~Eksin, ``Distributed networked learning with correlated data,'' in \emph{2020 59th IEEE Conference on Decision and Control (CDC)}.\hskip 1em plus 0.5em minus 0.4em\relax IEEE, 2020, pp. 5923--5928.

\bibitem{zhang2014regularization}
Y.~Zhang and D.-Y. Yeung, ``A regularization approach to learning task relationships in multitask learning,'' \emph{ACM Transactions on Knowledge Discovery from Data (TKDD)}, vol.~8, no.~3, pp. 1--31, 2014.

\bibitem{crawshaw2020multi}
M.~Crawshaw, ``Multi-task learning with deep neural networks: A survey,'' \emph{arXiv preprint arXiv:2009.09796}, 2020.

\bibitem{liu2017distributed}
S.~Liu, S.~J. Pan, and Q.~Ho, ``Distributed multi-task relationship learning,'' in \emph{Proceedings of the 23rd ACM SIGKDD International Conference on Knowledge Discovery and Data Mining}, 2017, pp. 937--946.

\bibitem{smith2017federated}
V.~Smith, C.-K. Chiang, M.~Sanjabi, and A.~S. Talwalkar, ``Federated multi-task learning,'' \emph{Advances in neural information processing systems}, vol.~30, 2017.

\bibitem{wang2016distributed}
J.~Wang, M.~Kolar, and N.~Srerbo, ``Distributed multi-task learning,'' in \emph{Artificial intelligence and statistics}.\hskip 1em plus 0.5em minus 0.4em\relax PMLR, 2016, pp. 751--760.

\bibitem{Nassif_2016}
R.~Nassif, C.~Richard, A.~Ferrari, and A.~H. Sayed, ``Multitask diffusion adaptation over asynchronous networks,'' \emph{IEEE Transactions on Signal Processing}, vol.~64, no.~11, pp. 2835--2850, 2016.

\bibitem{fan2016overview}
J.~Fan, Y.~Liao, and H.~Liu, ``An overview of the estimation of large covariance and precision matrices,'' \emph{The Econometrics Journal}, vol.~19, no.~1, pp. C1--C32, 2016.

\bibitem{ledoit2022power}
O.~Ledoit and M.~Wolf, ``The power of (non-) linear shrinking: A review and guide to covariance matrix estimation,'' \emph{Journal of Financial Econometrics}, vol.~20, no.~1, pp. 187--218, 2022.

\bibitem{precision}
W.~N. Van~Wieringen and C.~F. Peeters, ``Ridge estimation of inverse covariance matrices from high-dimensional data,'' \emph{Computational Statistics \& Data Analysis}, vol. 103, pp. 284--303, 2016.

\bibitem{Yu_2014}
Y.~Zhang and D.-Y. Yeung, ``A regularization approach to learning task relationships in multitask learning,'' \emph{ACM Transactions on Knowledge Discovery from Data (TKDD)}, vol.~8, no.~3, pp. 1--31, 2014.

\bibitem{Donsker}
M.~D. Donsker, \emph{An invariance principle for certain probability limit theorems}, 1951.

\bibitem{mortimore1989study}
P.~Mortimore, P.~Sammons, L.~Stoll, D.~Lewis, and R.~Ecob, ``A study of effective junior schools,'' \emph{International Journal of Educational Research}, vol.~13, no.~7, pp. 753--768, 1989.

\bibitem{oksendal2003stochastic}
B.~{O}ksendal, ``Stochastic differential equations,'' in \emph{Stochastic differential equations}.\hskip 1em plus 0.5em minus 0.4em\relax Springer, 2003, pp. 65--84.

\bibitem{sumchi}
A.~Ferrari, ``A note on sum and difference of correlated chi-squared variables,'' \emph{arXiv preprint arXiv:1906.09982}, 2019.

\bibitem{sumgamma}
Y.~Feng, M.~Wen, J.~Zhang, F.~Ji, and G.-x. Ning, ``Sum of arbitrarily correlated gamma random variables with unequal parameters and its application in wireless communications,'' in \emph{2016 International Conference on Computing, Networking and Communications (ICNC)}, 2016, pp. 1--5.

\bibitem{chiapproximated}
L.-L. Chuang and Y.-S. Shih, ``Approximated distributions of the weighted sum of correlated chi-squared random variables,'' \emph{Journal of Statistical Planning and Inference}, vol. 142, no.~2, pp. 457--472, 2012.

\bibitem{gammamultivariate}
J.~Zhang, M.~Matthaiou, G.~K. Karagiannidis, and L.~Dai, ``On the multivariate gamma--gamma distribution with arbitrary correlation and applications in wireless communications,'' \emph{IEEE Transactions on Vehicular Technology}, vol.~65, no.~5, pp. 3834--3840, 2015.

\bibitem{multiCLT}
W.~Whitt, ``An introduction to stochastic-process limits and their application to queues. internet supplement,'' 2002.

\end{thebibliography}
\bibliographystyle{IEEEtran}
 





\end{document}